\documentclass[twocolumn,showpacs,prl,amsmath,amstex,amssymb,citeautoscript,longbibliography]{revtex4-1}
\pdfoutput=1
\usepackage{natbib}
\usepackage[english]{babel}
\usepackage{letltxmacro}
\LetLtxMacro{\ORIGselectlanguage}{\selectlanguage}
\makeatletter
\DeclareRobustCommand{\selectlanguage}[1]{%
  \@ifundefined{alias@\string#1}
    {\ORIGselectlanguage{#1}}
    {\begingroup\edef\x{\endgroup
       \noexpand\ORIGselectlanguage{\@nameuse{alias@#1}}}\x}%
}
\newcommand{\definelanguagealias}[2]{%
  \@namedef{alias@#1}{#2}%
}
\makeatother
\definelanguagealias{en}{english}
\definelanguagealias{English}{english}
\usepackage{graphicx}
\usepackage{amsmath}
\usepackage{amsfonts}
\usepackage{amssymb}
\usepackage{bm}
\usepackage{color}
\usepackage[percent]{overpic}
\usepackage{soul} 
\usepackage{amssymb}
\usepackage{wasysym}
\usepackage{dsfont}

\usepackage{hyperref}
\hypersetup{
    bookmarks=false,         
    unicode=false,          
    pdftoolbar=false,        
    pdfmenubar=true,        
    pdffitwindow=false,     
    pdfstartview={FitH},    
    pdftitle={Power-Law Entanglement Spectrum in Many-Body Localized Phases},    
    pdfauthor={Maksym Serbyn, Alexios A. Michailidis, Dmitry A. Abanin, and Z. Papic},     
    pdfsubject={},   
    pdfcreator={},   
    pdfproducer={}, 
    pdfkeywords={many-body localization} {entanglement spectrum} {disordered systems}, 
    pdfnewwindow=true,      
    colorlinks=true,       
    linkcolor=black,          
    citecolor=blue,        
    filecolor=magenta,      
    urlcolor=blue           
}
\newcommand{\be}{\begin{equation}}
\newcommand{\ee}{\end{equation}}
\newcommand{\bea}{\begin{eqnarray}}
\newcommand{\eea}{\end{eqnarray}}

\newcommand{\la}{\langle}
\newcommand{\ra}{\rangle}

\renewcommand{\phi}{\varphi}
\renewcommand{\epsilon}{\varepsilon}

\newcommand{\corr}[1]{\langle{ #1}\rangle}

\newcommand{\ins}[1]{|{ #1}\rangle}
\newcommand{\out}[1]{\langle{ #1}|}
\newcommand{\kp}{{\kappa'}}
\newcommand{\kpp}{{\kappa}}

\begin{document}
\title{Power-Law Entanglement Spectrum in Many-Body Localized Phases
}
\author{Maksym Serbyn$^1$, Alexios A. Michailidis$^2$, Dmitry A. Abanin$^3$, and Z. Papi\'c$^4$}
\affiliation{$^1$ Department of Physics, University of California, Berkeley, California 94720, USA}
\affiliation{$^2$ School of Physics and Astronomy, University of Nottingham, Nottingham, NG7 2RD, UK}
\affiliation{$^3$ Department of Theoretical Physics, University of Geneva, 24 quai Ernest-Ansermet, 1211 Geneva, Switzerland}
\affiliation{$^4$ School of Physics and Astronomy, University of Leeds, Leeds, LS2 9JT, United Kingdom}
\date{\today}
\begin{abstract}
Entanglement spectrum of the reduced density matrix contains information beyond the von Neumann entropy and provides unique insights into exotic orders or critical behavior of quantum systems. Here we show that strongly-disordered systems in the many-body localized phase have universal power-law entanglement spectra, arising from  the presence of extensively many local integrals of motion.  The power-law entanglement spectrum distinguishes many-body localized systems from ergodic systems, as well as from ground states of gapped integrable models or systems in the vicinity of scale-invariant critical points. We confirm our results using large-scale exact diagonalization. In addition, we develop a matrix-product state algorithm which allows us to access the eigenstates of large systems close to the localization transition, and discuss general implications of our results for variational studies of highly excited eigenstates in many-body localized systems.
\end{abstract}
\maketitle

{\it Introduction.---}~Recently, much progress has been made towards understanding the mechanisms of ergodicity and its breakdown in an isolated quantum many-body system. Currently, two generic classes of many-body systems are known: ergodic (thermal) systems and many-body localized (MBL) systems ~\cite{Basko06,Mirlin05,OganesyanHuse,PalHuse}. An ergodic system is one that acts as a heat bath for its subsystems, and therefore thermalizes as a result of unitary evolution~\cite{DeutschETH,SrednickiETH,RigolNature}. By contrast, in MBL systems transport of energy is quenched by disorder, via a mechanism akin to the single-particle Anderson localization~\cite{Anderson58}. Nevertheless, MBL systems do reach stationary states~\cite{Serbyn14,Serbyn_14_Deer}, which are highly non-thermal due to the emergence of extensively many quasi-local integrals of motion (LIOMs)~\cite{Serbyn13-1,Huse13,Imbrie16}.

In addition to distinct dynamical properties,  ergodic and MBL systems are sharply distinguished by the microscopic nature of their eigenstates. This difference can be probed via quantum-information measures, such as entanglement entropy (EE). Given a pure quantum state $\psi$ of a many-body system ${\mathcal S} = {\cal L}\cup{\cal R}$, consisting of two subsystems ${\mathcal L}$ and ${\cal R}$, the EE is defined as $S=-\sum_i^D \lambda_i \ln \lambda_i$, where $\{ \lambda_i \}, \, i=1,...,D$, are the eigenvalues of the reduced density matrix $\hat \rho_{\mathcal R}={\rm Tr}_{{\mathcal L}} |\psi\ra \la \psi|$, and $D$ is the dimensionality of the Hilbert space of ${\mathcal R}$. The EE of highly-excited eigenstates of thermal systems, which obey the ``Eigenstate Thermalization Hypothesis"~\cite{DeutschETH,SrednickiETH,RigolNature}, is known to generically scale as the number of degrees of freedom in $\cal R$ (``volume law").
On the other hand, in an MBL system the EE of nearly all eigenstates  obeys the ``area law" \cite{Serbyn13-1,Huse13,Bauer13}. This weaker scaling of EE makes MBL systems reminiscent of ground states of gapped systems~\cite{Hastings07}. 

EE, while providing a quantitative measure of entanglement in a many-body state, contains no information about how it is created or how different degrees of freedom are entangled with each other. Therefore, to gain a better understanding of the structure of MBL and ergodic states, we study the ``entanglement spectrum" (ES) \cite{Haldane08}, i.e., the full eigenspectrum of the reduced density matrix,  $\{ \lambda_i \}$. The ES has been extensively studied in free fermion~\cite{Peschel2,*Peschel} and critical systems~\cite{Calabrese}. A particular advantage of the ES is that it can characterize and classify exotic quantum orders that cannot be described by symmetry-breaking \cite{Haldane08, PollmannTurnerBergOshikawa, Fidkowski, YaoQi}. 

\begin{figure}[htb]
\centering
\setlength{\unitlength}{\columnwidth}
\begin{picture}(0,0)
\put(0.38, 0.42){\includegraphics[width=0.48\columnwidth]{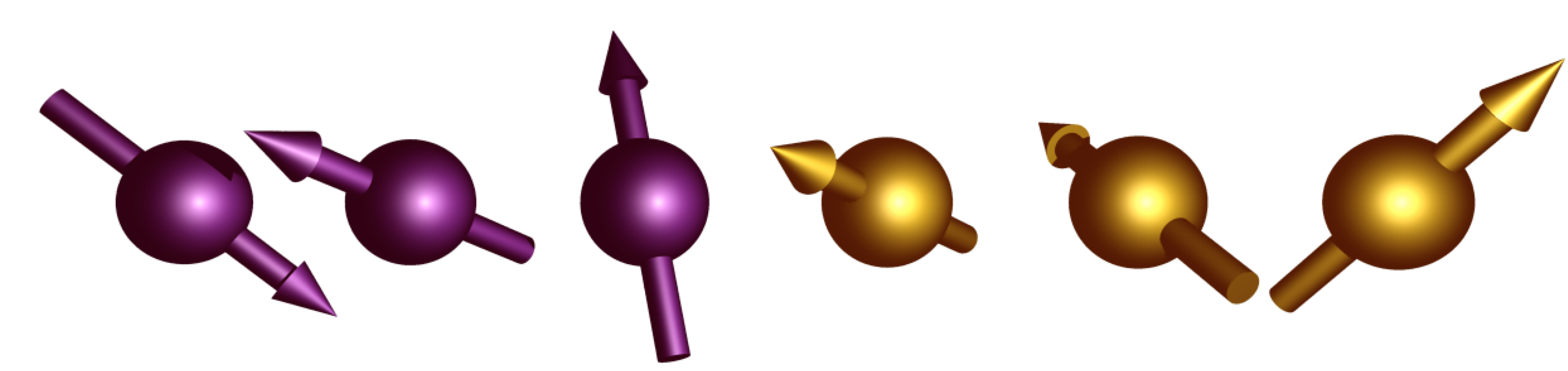}}
\put(0.5, 0.39){$\cal L$}
\put(0.72, 0.39){$\cal R$}
\end{picture}
\includegraphics[width=0.96\columnwidth]{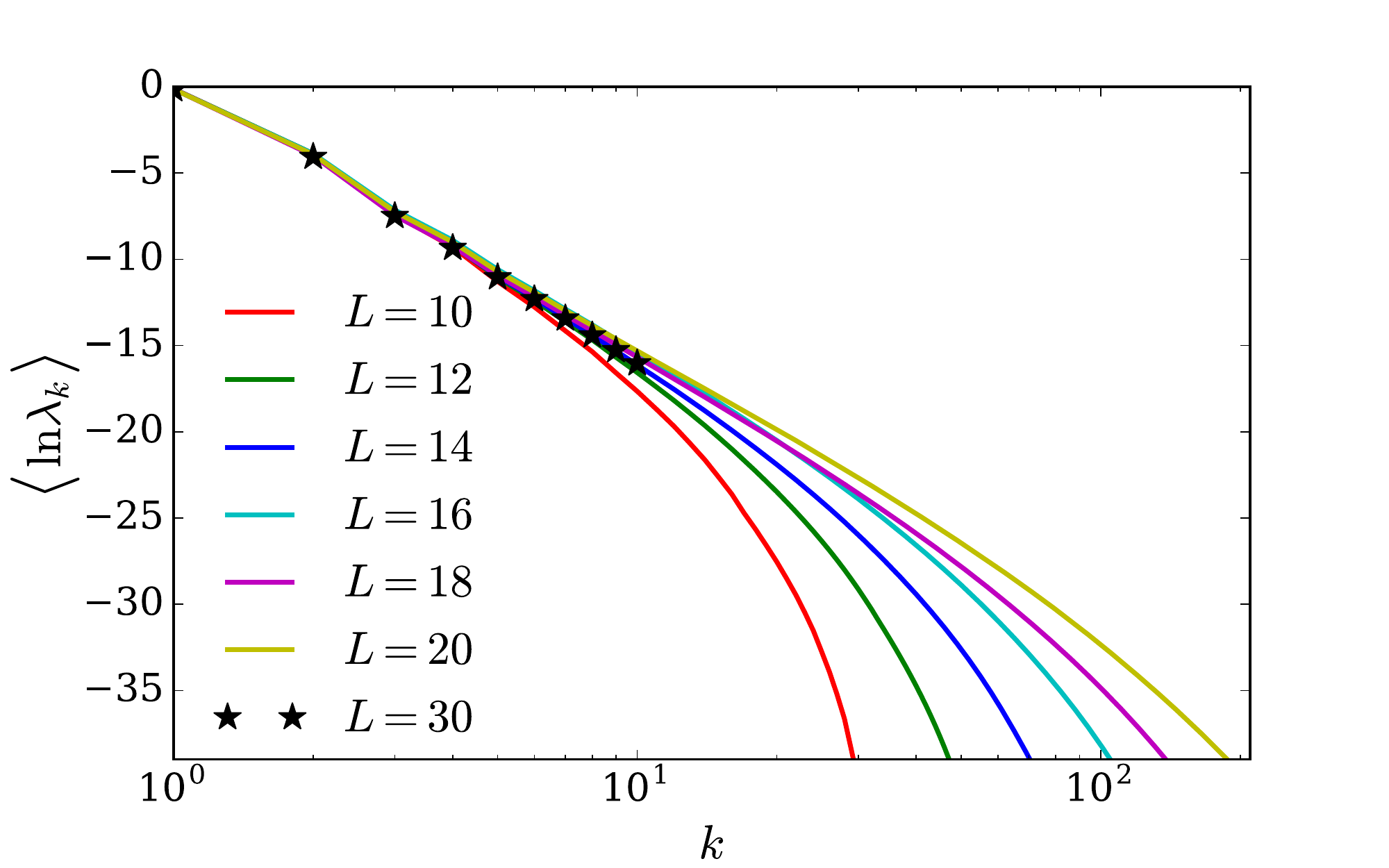}
\caption{ \label{Fig:entspec} (Color online) ES of the highly-excited eigenstates of XXZ spin chain with disorder strength $W=5$. The spectrum has a power law form in the MBL phase and in the vicinity of the delocalization transition.}
\end{figure}

In this paper, we obtain a more complete understanding of the eigenstate entanglement properties in the MBL phase and in the vicinity of the delocalization transition. We demonstrate that the ES in the MBL phase has a universal \emph{power-law structure}, whose exponent is proportional to the many-body localization length [Fig.~\ref{Fig:entspec}]. This structure results from the fact that the ES probes the correlations at the boundary between the subsystems $\cal L$ and $\cal R$, and due to the existence of an extensive number of local operators that commute with the Hamiltonian in the MBL phase~\cite{Serbyn13-1,Huse13,Imbrie16}. Thus, the universal power-law distinguishes MBL systems from ergodic systems where the ES obeys the Marchenko-Pastur distribution \cite{MarcenkoPastur,Chamon15}. Moreover, the power-law spectrum reveals a difference between MBL systems and ground states of gapped integrable models \cite{Alba12} or free systems in the vicinity of scale-invariant critical points \cite{Calabrese,Cho16}, where the ES typically decays faster than power law \cite{Peschel2, Calabrese, Yasuhiro99, Franchini2011}.

In addition to providing new insights into the properties of MBL systems, the ES is of crucial importance for the matrix-product state (MPS) optimization algorithms such as the ``density matrix renormalization group" (DMRG) \cite{White}. While in principle the MPS naturally encode the eigenstates of MBL phase due to the area-law entropy, practical realizations of the efficient optimization algorithms are an active area of research~\cite{Pekker15,Sheng15,Pollmann15-1,Pollmann15,Karrasch15}. In this work we develop an MPS optimization to target the highly-excited states of a disordered XXZ chain in 1D, and use the power-law ES as a sensitive benchmark of its accuracy in large systems up to $L=30$ spins.  Our analytic results for the ES allow us to put bounds on the bond dimension, and demonstrate the feasibility of the DMRG calculation of highly excited states in close proximity to the delocalization transition. 

Our work complements the recent work by Yang \emph{et al.}~\cite{Chamon15} and Geraedts \emph{et al.}~\cite{Rahul16}, who studied the ES distribution and level statistics in ergodic and MBL phases, and Monthus \cite{Monthus}, who derived the scaling of Renyi entropies in the MBL phase in first-order perturbation in the coupling between the subsystems.

{\it Model.---}We consider a standard  model of MBL -- a XXZ spin-$1/2$ chain of $L$ spins with a random $z-$field~\cite{PalHuse}:
\be\label{eq:model}
H=\frac12\sum_{i=1}^{L-1}\left[J_x( \sigma_i^x \sigma_{i+1}^x+\sigma_i^y \sigma_{i+1}^y )+J_z\sigma_i^z \sigma_{i+1}^z\right]+\sum_{i=1}^L h_i \sigma_i^z, 
\ee
where  $h_i\in [-W;W]$ are independent, uniform random numbers, and $\sigma^\alpha$ are the Pauli matrices. We choose open boundary conditions and assume a bipartition that separates the system into equal $\cal L$ and $\cal R$ parts [Fig~\ref{Fig:entspec}, inset].

The model (\ref{eq:model}) has been extensively studied and is believed to capture all essential properties of the MBL phase and the localization transition. For example, it is known that the model supports an MBL phase at strong disorder, an ergodic phase at weaker disorder, and an integrable point at zero disorder. For $J_x=J_z=1$, the transition between the two phases was estimated to be at $W_c\approx 3.5$ based on a variety of probes, for example the level statistics~\cite{PalHuse,Alet14,Serbyn16}, fluctuations of EE~\cite{Kjall14}, and the statistics of the matrix elements of local operators~\cite{Serbyn15}. 

{\it Power-law entanglement spectrum.---} Before discussing numerical results for the model~(\ref{eq:model}), we infer the general properties of the ES in the MBL phase from the existence of LIOMs~\cite{Serbyn13-1,Huse13,Imbrie16,ScardicchioLIOM}.  In the ``fully" MBL phase (i.e., when there is no mobility edge in the spectrum~\cite{Alet14,Serbyn15}), there exists a quasi-local unitary transformation which diagonalizes the Hamiltonian by rotating the physical spins $\sigma_i$ into the exactly conserved LIOMs  $\tau_i$. The latter form a complete basis of the Hilbert space, and any many-body eigenstate is a simultaneous eigenstate of all $\tau^z_i$, $i=1,\ldots L$.

 Let us expand a given eigenstate $\ins{I}$ over the complete basis formed by tensor product of eigenstates in $\cal L$ and~$\cal R$:
\begin{equation}\label{Eq:I-exp}
 \ins{I}
 =
\sum_{\{\mu\}_{\cal L} ,\{\tau\}_{\cal R} } 
C_{\{\mu\}_{\cal L} \{\tau\}_{\cal R}} 
\ins{\{\mu\}_{\cal L}}
\otimes
\ins{\{\tau\}_{\cal R}}.
\end{equation}
In the MBL phase, the values of  LIOMs  in $\cal L$ or $\cal R$, $\{\mu\}_{\cal L}$ and $\{\tau\}_{\cal R}$ respectively, label the basis vectors. In this basis, 
the reduced density matrix of the state~(\ref{Eq:I-exp}) for $\mathcal{R}$ subsystem reads
$
 \out{\{\chi\}_{\cal R}}
 \hat \rho_\mathcal{R} \ins{\{\tau\}_{\cal R}}
 =
 \sum_{\{\mu\}_{\cal L}} 
 C^*_{\{\mu\}_{\cal L} \{\chi\}_{\cal R}} 
 C_{\{\mu\}_{\cal L} \{\tau\}_{\cal R}} 
 $,
where the sum over all configurations of the $\cal L$ subsystem arises from partial trace. We rewrite this matrix as
$
 \hat \rho_\mathcal{R}
 =
 \sum_{\{\mu\}_{\cal L}} \ins{\psi_{\{\mu\}_{\cal L}}}\out{\psi_{\{\mu\}_{\cal L}}}.
 $
 The vectors $\ins{\psi_{\{\mu\}_{\cal L}}}$ are given by the coefficients in Eq.~(\ref{Eq:I-exp}): 
\begin{equation}\label{Eq:psi-def}
  \ins{\psi_{\{\mu\}_{\cal L}}}
  =
  \left(C_{\{\mu\}_{\cal L} \{\tau_1\}_{\cal R}},C_{\{\mu\}_{\cal L} \{\tau_2\}_{\cal R}}
  \ldots,
 C_{\{\mu\}_{\cal L} \{\tau_{D_{\cal R}}\}_{\cal R}}
  \right)^T,
\end{equation}
where each of $D_{\cal R}=2^{L_{\cal R}}$ components  is labeled by the different configurations of LIOMs in $\cal R$.

Deep in the MBL phase, to the order $\mathcal{O}(1)$, an eigenstate $\ins{I}$ of the full system is a product state of certain eigenstates of $\cal L$ and $\cal R$ subsystems. Let us define the LIOMs such that these eigenstates are labelled by configurations with all effective spins pointing up,  $\tau^z_i \ins{I} = \ins{I}$ for all $i$. Then, in the expansion~(\ref{Eq:I-exp}), the largest coefficient is $|C_{\{\mu\}_{\cal L} \{\tau\}_{\cal R}}|=c_0$, with both $\{\tau\}_{\cal R}$ and $\{\mu\}_{\cal L}=\uparrow\uparrow\ldots\uparrow$. The typical value of a coefficient with some of the LIOMs flipped is suppressed as 
\begin{equation}\label{Eq:C-supp}
|C_{\{ \uparrow \ldots\uparrow
\underbrace{\scriptstyle
\downarrow\downarrow\uparrow \}_{\cal L} \{ \uparrow \uparrow  \downarrow 
}_{r}
\uparrow\ldots \uparrow  \}_{\cal R}}| \approx c_0 e^{-\kpp r},
\end{equation}
where $r$ specifies  the ``radius of the disturbance" (RoD) of effective spins near the entanglement cut, and $\kpp$ is the inverse characteristic (many-body) localization length.  Note that~$\kpp$ may fluctuate depending on disorder pattern, and should not be taken as a direct analogue of the single-particle localization length as it does not diverge at the transition~\cite{Serbyn15}.

If we order the basis in $\cal R$ according to the RoD, the exponential suppression~(\ref{Eq:C-supp}) implies that (i) all terms in  $\ins{\psi_{\{\mu\}_{\cal L}}}$ are suppressed as $e^{-\kpp r_{\cal L}}$, where $r_{\cal L}$ is the RoD in the left subsystem; (ii) components of  $\ins{\psi_{\{\mu\}_{\cal L}}}$ are ordered according to their magnitude, so that the first term (corresponding to no spin flips in $\cal R$) is of order one, the term with one spin flip is of the order $e^{-\kpp}$, etc. Denoting $a =e^{-\kpp}$, a typical  $\ins{\psi_{\{\mu\}_{\cal L}}}$ is:
\begin{multline}\label{Eq:psi-mag}
  \ins{\psi_{\{\mu\}_{\cal L}}}
=
a^{r_{\cal L}}
(\alpha_1;\ \alpha_2 a;\ \alpha_3 a^2, \alpha_4 a^2;\ \alpha_5 a^3,\ldots ,\alpha_8 a^3;\ \\
\ldots;\ \alpha_{1+D_{\cal R}/2}  a^{L_{\cal R}},\ldots, \alpha_{D_{\cal R}}  a^{L_{\cal R}} )^T,
\end{multline}
where all $ |\alpha_i|$ are assumed to be of order one, and we separated the blocks corresponding to the value of RoD $r_{\cal R}=0,1,2,\ldots, L_{\cal R}$ by semicolons.

If different  vectors $\ins{\psi_{\{\mu\}_{\cal L}}}$ in Eq.~(\ref{Eq:psi-def}) were mutually orthogonal, their norm $\corr{\psi_{\{\mu\}_{\cal L}}|\psi_{\{\mu\}_{\cal L}}}\propto e^{-2 \kpp r_{\cal L}}$ would give the eigenvalues of $\hat\rho_\mathcal{R}$, and hence the ES. In the Supplemental Material~\cite{SOM} we demonstrate it is possible to perturbatively orthogonalize the vectors $\ins{\psi_{\{\mu\}_{\cal L}}}$ deep in the MBL phase where $e^{-\kpp}\ll 1$. This process results in the eigenvalues labeled by the RoD $r$:
\begin{equation}\label{Eq:ld-order}
  \lambda^{(r)}_k 
  =
  \lambda_{ \uparrow\ldots\uparrow\underbrace{\scriptstyle
  \downarrow\ldots \downarrow }_r} \propto e^{-4\kpp r}, 
\end{equation}
where $k= 2^{r-1}+1,\ldots, 2^r$ labels $2^{r-1}$ different eigenvalues in the block corresponding to RoD $r$. An extra factor of 2 in the exponent in Eq.~(\ref{Eq:ld-order}) compared to the norm of corresponding $\ins{\psi_{\{\mu\}_{\cal L}}}$ arises from the fact that all components in $\ins{\psi_{\{\mu\}_{\cal L}}}$, corresponding to blocks with RoD less than $r$, are cancelled in the process of orthogonalization~\cite{SOM}. Intuitively, this means that the processes, which  contribute to eigenvalues with RoD equal to $r_{\cal L}$ in the $\cal L$ subsystem, flip the same number of spins in the $\cal R$ subsystem. 

One can view the RoD $r$, or equivalently the typical number of spin flips, as an effective ``quantum number'' underlying the structure of the ES. This is analogous to, e.g., the subsystem's momentum perpendicular to the entanglement cut (which also labels the edge states if a system has topological order); similar structure for the XXZ ground state was pointed in Ref.~\cite{Alba-12}.

The hierarchical structure of the reduced density matrix implies a \emph{power-law} structure of the  ES as a function of $k$.  Indeed, expressing $r$ as $r \approx \ln k/\ln 2$, and using Eq.~(\ref{Eq:ld-order}), we find the typical value of $\lambda_k$
\begin{equation}\label{Eq:power-law}
\lambda_k
\propto 
\frac{1}{k^\gamma},
\qquad
\gamma
\simeq\frac{4\kpp}{ \ln2}.
\end{equation}
to decay as a power law with exponent set by $\kpp$~\cite{Note1}.

In addition, we can also understand the finite-size effects in the ES. The power-law holds until the very last block, for which $r = L_{\cal L}$. The average value of $\lambda_k$ for $k\gtrsim 2^{L_{\cal L}-1}$ will deviate from the simple power-law form~(\ref{Eq:power-law}). Instead, $\log\lambda_k$ will be given by the order statistics of the Gaussian distribution arising from log-normal statistics of the coefficients~(\ref{Eq:C-supp})~\cite{Luca13,Serbyn15} in the MBL phase, which describes accurately the tail of the ES {as we demonstrate in~\cite{SOM}}. 

\begin{figure}[t]
\begin{center}
\includegraphics[width=0.99\columnwidth]{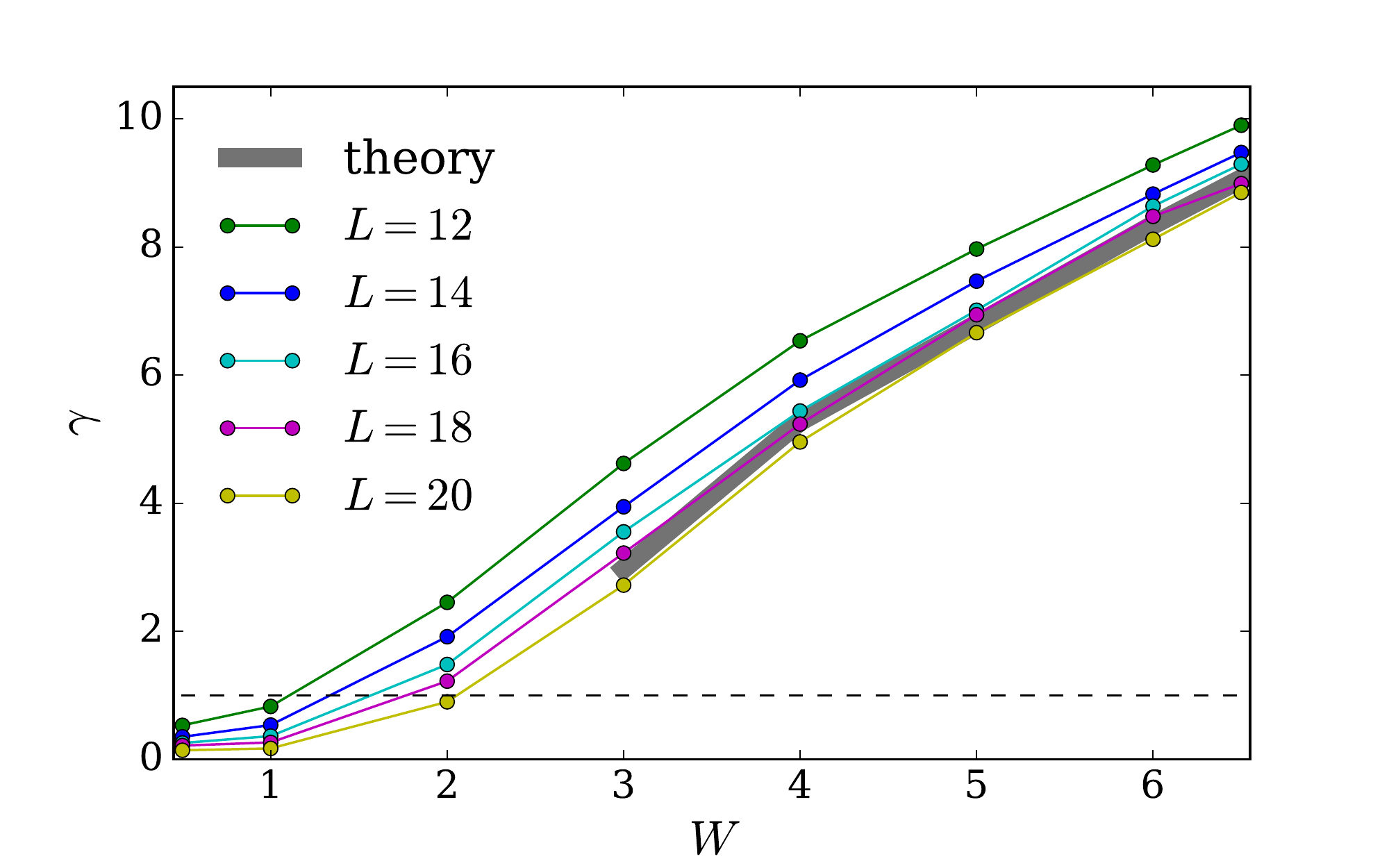}
\caption{ \label{Fig:gamma} (Color online) Power-law exponent $\gamma$, extracted from the fit of the ES, $\langle\ln\lambda_k\rangle$, increases with disorder $W$. Theoretical prediction refers to $\gamma$ extracted from the scaling of the matrix elements in Ref.~\cite{Serbyn15}.}
\end{center}
\end{figure}

\begin{figure*}
\begin{center}
\includegraphics[width=0.659\columnwidth]{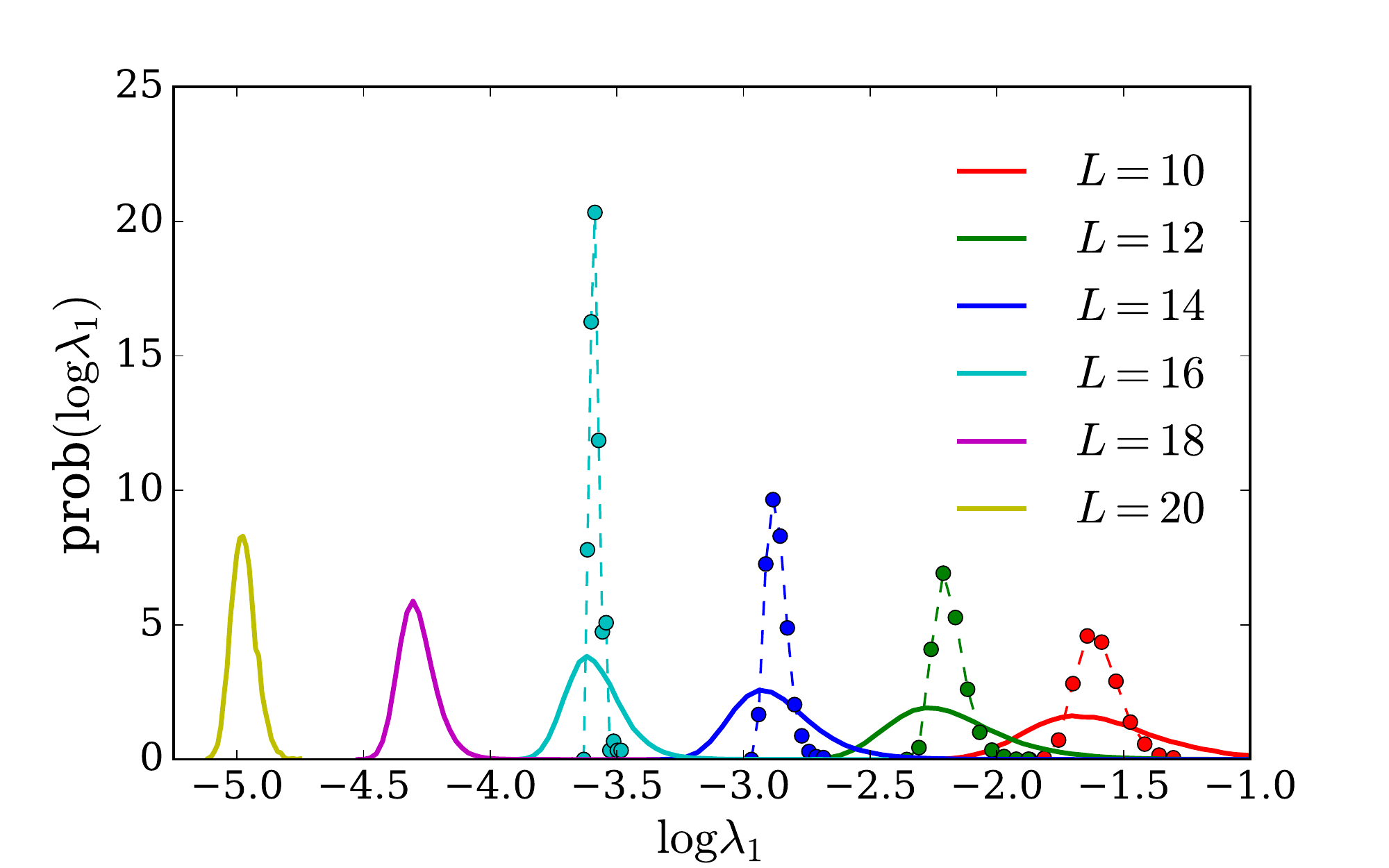}
\includegraphics[width=0.659\columnwidth]{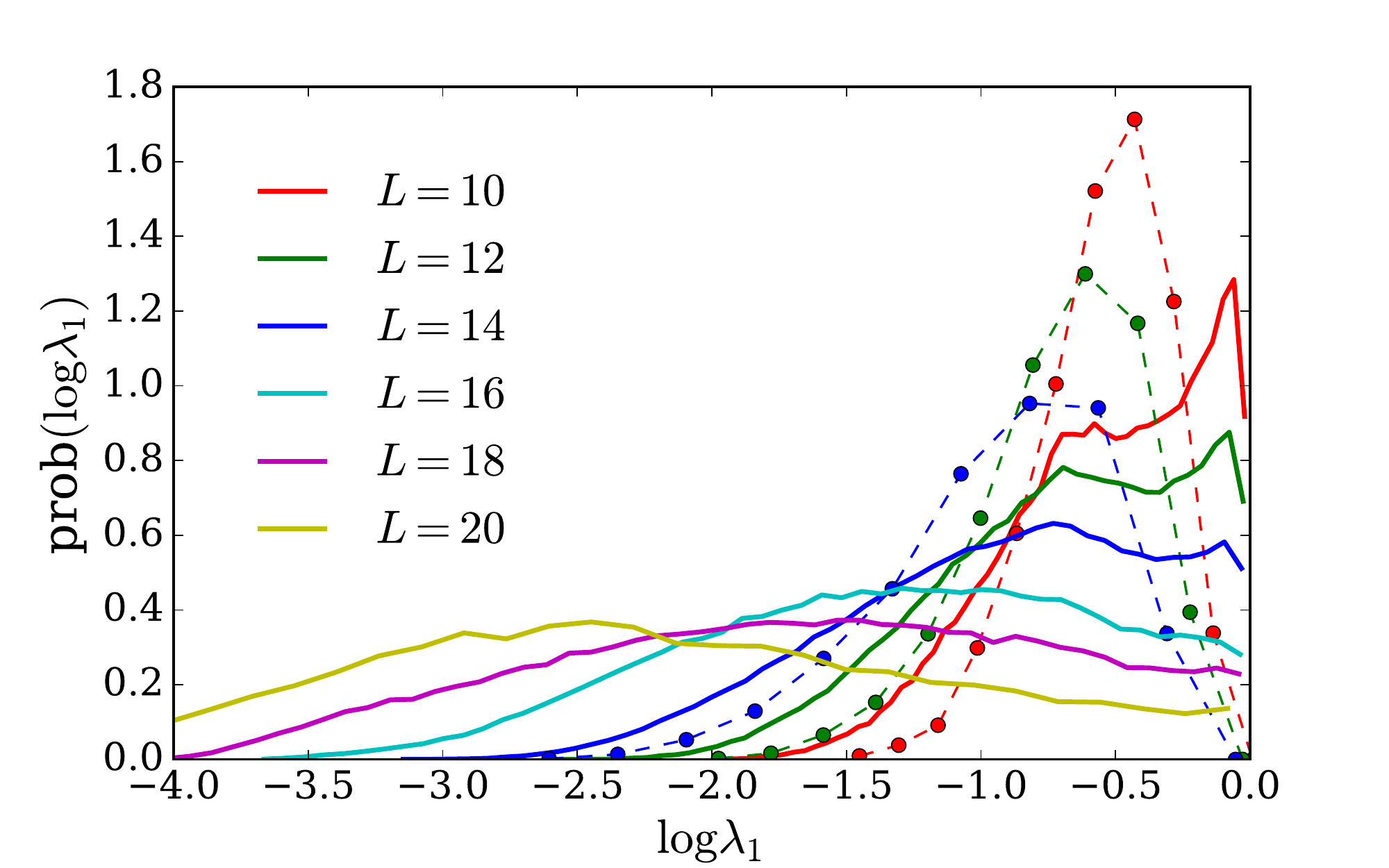}
\includegraphics[width=0.659\columnwidth]{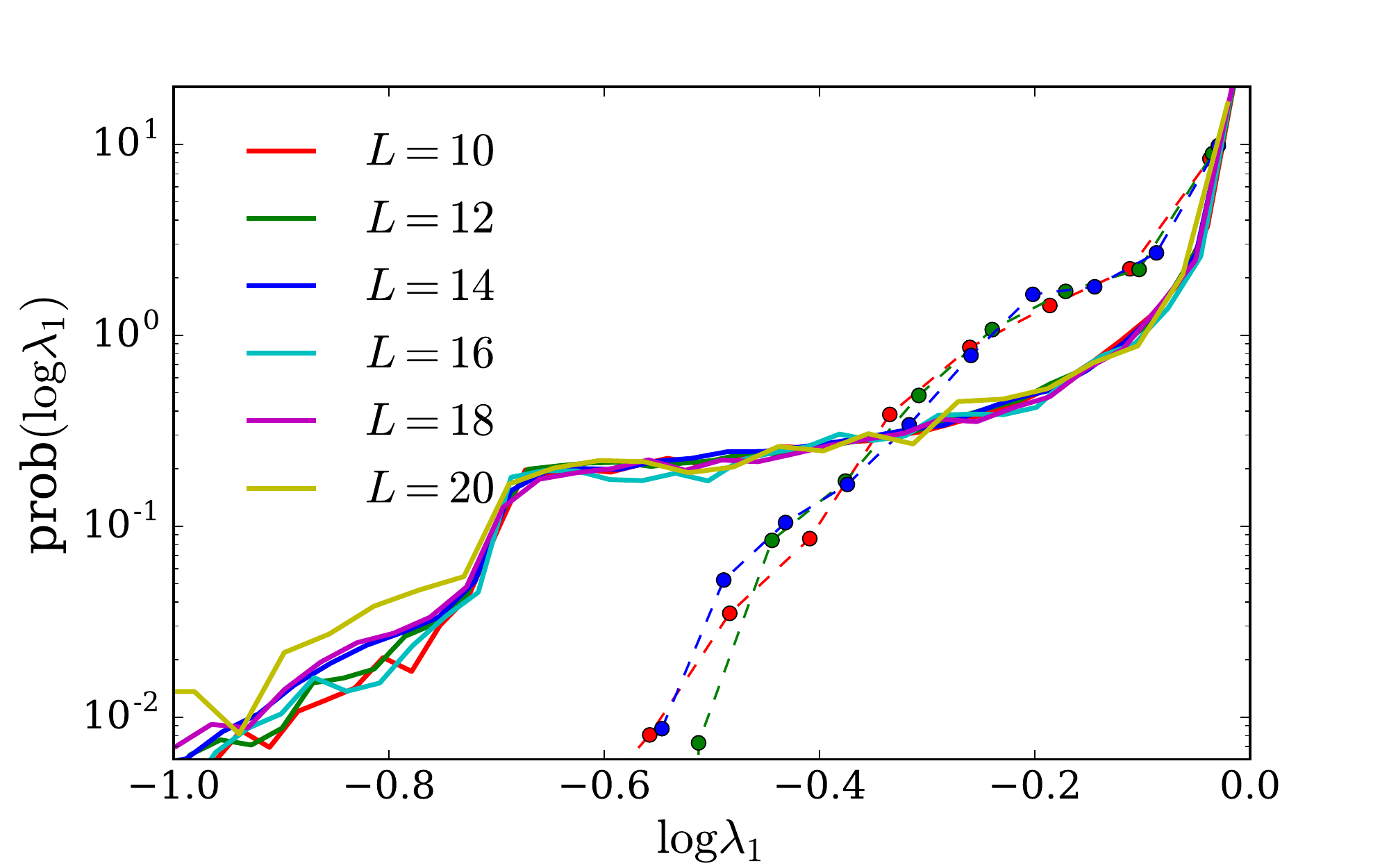}
\caption{ \label{Fig:lmax} (Color online) Distribution of $\lambda_1$ across the MBL transition for different $L$. Solid lines indicate full distribution, dashed lines show distribution of $\lambda_1$ between \emph{different disorder realizations}. Disorder strength is $W=0.5$ (left), $W=2$ (middle), $W=6.5$ (right).   }
\end{center}
\end{figure*}
{\it Numerical results.---}To study the ES numerically in the XXZ chain~(\ref{eq:model}), we use: (i) full exact diagonalization (ED) for $L=10, 12, 14$ spins, (ii) ``shift and invert" algorithm (SI)~\cite{SLEPc} for $L=16, 18, 20$, and (iii) a new implementation of the MPS variational optimization  for larger $L$ (below we present data for $L=30$).  Our MPS algorithm combines the advantage of SI spectral transformation, which ensures low energy variance and hence the purity of eigenstates, with a fast conjugate-gradient linear solver. The MPS optimization converges efficiently when the bond dimension $\chi_{max}$ is such that $\ln(\chi_{max}) \gg S$, where $S$ is the maximum entropy for all partitions of the chain. Using ITensor libraries~\cite{Note2} 
with conserved $U(1)$ symmetry and an iterative local scheme, we can reach $\chi_{max}\approx 500$, thus capturing a big part of the ES without finite-bond effects \cite{SOM}.

Fig.~\ref{Fig:entspec}  illustrates the log-averaged ES, {defined as $\{\langle\ln\lambda_k\rangle\}$, where $\lambda_k$ are ordered from largest to smallest magnitude, and brackets denote averaging over disorder}, as a function of the eigenvalue number $k$, for various system sizes $L$. Consistent with our expectations~(\ref{Eq:power-law}), in the MBL phase ($W=5$) the ES exhibits clear power-law behavior. In all cases, we  target the eigenstates close to energy $E=0$, which is roughly in the middle of the many-body band. The data is averaged over a few thousand disorder realizations for $L\leq 16$, and over a few hundred realizations for for $L=18, 20$. For $L=30$, we used $\chi_{max}=200$ and 1000 disorder realizations. 

Note that, while we find excellent agreement between ED and MPS results for the few largest Schmidt eigenvalues, the lowest Schmidt values obtained by MPS lie slightly below the ED data for $L=20$. This is an artefact of our fixed bond dimension $\chi_{max}=200$, which bounds the slope of the ES through its effect on the smallest Schmidt values. For the given $\chi_{max}$, we expect the MPS slope to be close to the  
exact slope of the system $L\sim 2\log_2\chi_{max}$, or $L\sim 14$ in our case (as  Fig. \ref{Fig:entspec} confirms). Note that this is a subtle effect which only affects the tail of the ES, while the quantities such as energy or entropy are converged to machine precision~\cite{SOM}.

Next, we study the behavior of the exponent $\gamma$ extracted from the power-law fit of $\lambda_k$ for small $k$. The exponent $\gamma$ always decreases with system size $L$, as can be seen in Fig.~\ref{Fig:gamma}. In the MBL phase we expect the exponent to saturate to a finite value, which is set by $\kpp$ governing the coefficients in Eq.~(\ref{Eq:I-exp}). To the leading order in perturbation, the relevant coefficients are a product of matrix elements in $\cal L$ and $\cal R$ over the energy denominator $\Delta$,
$
C_{\{\mu\}_{\cal L}\{\tau\}_{\cal R}}\approx \corr{\{\mu\}_{\cal L}|S^\alpha_{L/2}|\{\uparrow\}_{\cal L}}\corr{\{\tau\}_{\cal R}|S^\alpha_{L/2+1}|\{\uparrow\}_{\cal R}}/\Delta. 
$
The typical value of $C$ corresponds to when both matrix elements flip a similar number of spins. The value of   $\kpp$ can be approximated as $2{\kpp} \approx 2{\kp}+\ln 2$, where $\kp$ governs the decay of the many-body analogue of the Thouless conductance $\cal G$, introduced in Ref.~\cite{Serbyn15}. Fig.~\ref{Fig:gamma} shows that this theoretical expectation describes accurately the power-law coefficient $\gamma$ at sufficiently large $L$. Note that in the ergodic phase ($W=0.5$), the power $\gamma$ is not well defined as the ES obeys a qualitatively different Marchenko-Pastur distribution~\cite{Chamon15}. 

\begin{figure}[b]
\begin{center}
\includegraphics[width=0.99\columnwidth]{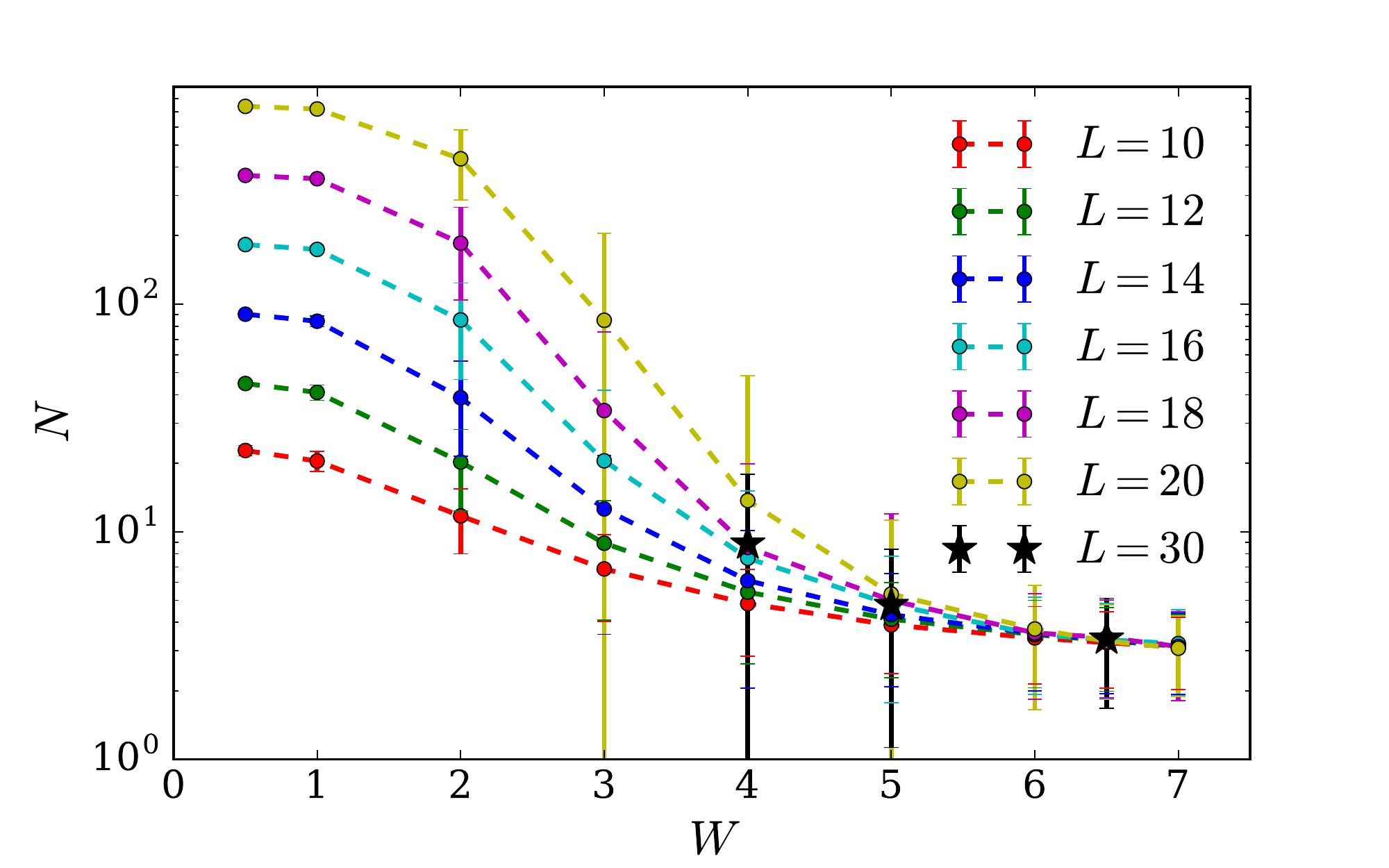}
\caption{ \label{Fig:N} (Color online) Number of singular values required to reproduce the entanglement entropy with fixed precision ($99\%$) decreases with disorder  strength $W$ and saturates at strong disorder. The bars represent statistical fluctuations, which are most pronounced near the transition.}
\end{center}
\end{figure}

{\it Sample-to-sample fluctuations.---}So far we discussed behavior of the log-averaged ES. Now we consider  the distribution of the ES for different disorder realizations in order to understand whether the ES statistics is dominated by  sample-to-sample fluctuations, or rather the fluctuations between different eigenstates in a single disorder realization. 

The distribution of the largest ES eigenvalue, $\lambda_1$, and its dependence on $L$ is illustrated in Fig.~\ref{Fig:lmax}. In the ergodic phase ($W=0.5$),  the center of the distribution of $\lambda_1$ shifts to smaller values \cite{MarcenkoPastur}, and becomes increasingly narrower with increasing $L$, reflecting the fact that all eigenstates become typical.  On the other hand, deep in the MBL phase, the distribution of $\lambda_1$ depends very weakly on $L$, as expected (Fig.~\ref{Fig:lmax}, $W=6.5$). Moreover, the peak in $\lambda_1$ is very close to one, indicating that eigenstates in the MBL phase are  well-approximated by product states. 

Finally, near the transition (Fig.~\ref{Fig:lmax}, $W=2$), the distribution of $\lambda_1$ becomes very broad, reflecting the fact that certain disorder realizations are insulating, while others are metallic.  Using ED data, we also average the leading eigenvalue over a window of eigenstates from a given disorder realization, and bin the resulting $\corr{\log\lambda_1}_\text{e.s.}$. Distribution of $\corr{\log\lambda_1}_\text{e.s.}$, shown by dashed lines in Fig.~\ref{Fig:lmax} (middle),  has the same width as the full distribution of~$\lambda_1$. This implies that the broad distribution of $\lambda_1$ near the MBL transition originates from sample-to-sample fluctuations, {provided that one fixes the position of the entanglement cut. Note, that recently large entanglement fluctuations w.r.t.\ the position of the cut within the same disorder realization were reported~\cite{Clark16}.}

{\it Discussion.---}
We demonstrated a power-law decaying ES in MBL states, which is in sharp contrast with both thermal systems, whose ES is ``flat'' \cite{Chamon15}, and ground states of gapped free or integrable models, whose ES decays faster than power law~\cite{Peschel2, Calabrese,Yasuhiro99,Franchini2011}. We used this distinct feature of MBL systems to perform highly sensitive benchmarks of our MPS algorithm. Using our algorithm, we obtained eigenstates of large systems  at disorder $W=4$, which is closer to the MBL transition than previously reported~\cite{Pekker15,Sheng15,Pollmann15-1,Pollmann15,Karrasch15}.

The power-law ES  implies that finite-size effects from the truncation of the ES -- a standard procedure in the MPS-like algorithms -- typically decay algebraically with $L$. In Fig.~\ref{Fig:N} we show the estimate for the MPS bond dimension required to reproduce the exact EE within $1\%$. While at weak disorder the estimate grows exponentially with $L$, in the MBL phase it saturates to a constant in a power-law fashion (not shown). Note that even at disorder $W\geq 3$ we have a value of $N\lesssim 100$, hence explaining the success of our MPS algorithm, and suggesting it is feasible to push such algorithms even closer to the MBL transition. 

Finally, the organization of the ES according to the number of spin flips by the boundary perturbation may have a number of consequences beyond the power-law structure of ES. In particular, it would be interesting to explore its significance for the ES level statistics studied in Ref.~\cite{Rahul16}, and use it to extract  $\kpp$ and other information about LIOMs from individual eigenstates.

{\it Acknowledgments.---}We thank M.~Stoudenmire and C.~Turner for useful discussions. M.S. was supported by Gordon and Betty Moore Foundation's EPiQS Initiative through Grant GBMF4307. This research was supported in part by the National Science Foundation under Grant No. NSF PHY11-25915, and by Swiss National Science Foundation and Alfred Sloan Foundation (DA). This work made use of the facilities of N8 HPC Centre of Excellence, provided and funded by the N8 consortium and EPSRC (Grant No.EP/K000225/1). The Centre is co-ordinated by the Universities of Leeds and Manchester.

%

\clearpage
\clearpage
\pagebreak
\onecolumngrid
\begin{center}
\textbf{\large Supplemental Online Material for ``Power-Law Entanglement Spectrum in Many-Body Localized Phases''}\\[5pt]
\begin{quote}
{\small In this supplementary material we present additional details on the derivation of the power-law entanglement spectrum in the MBL phase. Moreover, we show numerical results for the distribution of eigenvalues of the reduced density matrix and for the tails of the entanglement spectrum. In the second part, we describe technical details behind the implementation of the MPS algorithm which was used to study large chains with $L=30$ spins.}\\[20pt]
\end{quote}
\end{center}
\setcounter{equation}{0}
\setcounter{figure}{0}
\setcounter{table}{0}
\setcounter{page}{1}
\setcounter{section}{0}
\makeatletter
\renewcommand{\theequation}{S\arabic{equation}}
\renewcommand{\thefigure}{S\arabic{figure}}
\renewcommand{\thesection}{S\Roman{section}}
\renewcommand{\thepage}{S\arabic{page}}

\twocolumngrid

\section{A. Singular values of the density matrix}
\subsection{A.1 Derivation of the power-law entanglement spectrum}
In the main text we demonstrated that the density matrix can be written as:
\begin{equation}\label{Eq:rho-exp-vec2-SM}
 \hat \rho
 =
\sum_{\{\mu\}_{\cal L}} \ins{\psi_{\{\mu\}_{\cal L}}}\out{\psi_{\{\mu\}_{\cal L}}},
\end{equation}
where vectors $\ins{\psi_{\{\mu\}_{\cal L}}}$ are non-orthogonal, and have a certain hierarchy in the MBL phase. In particular, the magnitude of corresponding vector $\ins{\psi_{\{\mu\}_{\cal L}}}$,
\begin{equation}\label{Eq:norm}
 \tilde \lambda_{\{\mu\}_{\cal L}} = \corr{\psi_{\{\mu\}_{\cal L}}|\psi_{\{\mu\}_{\cal L}}}
\end{equation}
is suppressed with the number of spin flips in $\{\mu\}_{\cal L}$ as:
\begin{equation}\label{Eq:ld-hier-SM}
\tilde \lambda_{\uparrow\ldots\uparrow\underbrace{\scriptstyle
 \uparrow\ldots \downarrow
 }_r
 } \propto  e^{-2\kpp r}. 
\end{equation}
Defining $a=e^{-\kpp}\ll 1$, we get the following hierarchy of different $\tilde\lambda_i$, organized according to the region of disturbance $r$ which runs from 0 to the number of spins in the $\cal L$ subsystem:
\begin{subequations}\label{Eq:hier}
\begin{eqnarray}
 r=0: &\qquad\qquad& \tilde \lambda_1  \propto 1,
 \\\label{Eq:r1}
 r=1: &\qquad&  \tilde \lambda_2 \propto a^2,
  \\
 r=2: &\qquad&  \tilde \lambda_{3,4} \propto a^4, 
  \\
 r=3: &\qquad&   \tilde\lambda_{5,\ldots,8}\propto a^6, 
 \\
   &\ldots&
  \\
  r=L_{\cal L}: &\qquad&     \tilde \lambda_{2^{r-1}+1,\ldots, 2^r}\propto a^{2r}.
\end{eqnarray}
  \end{subequations}

If different vectors $\ins{\psi_i}$ (below we switch to labelling by index rather than spin configuration) were orthogonal to each other, then the $\tilde \lambda_i$  would give us the set of singular values of $\hat \rho$. Let us demonstrate that we can perturbatively orthogonalize all $\ins{\psi_i}$, and demonstrate that the eigenvalues of $\rho$ have similar hierarchical structure to the one in Eq.~(\ref{Eq:hier}). 

\begin{figure}[b]
\begin{center}
\includegraphics[width=0.99\columnwidth]{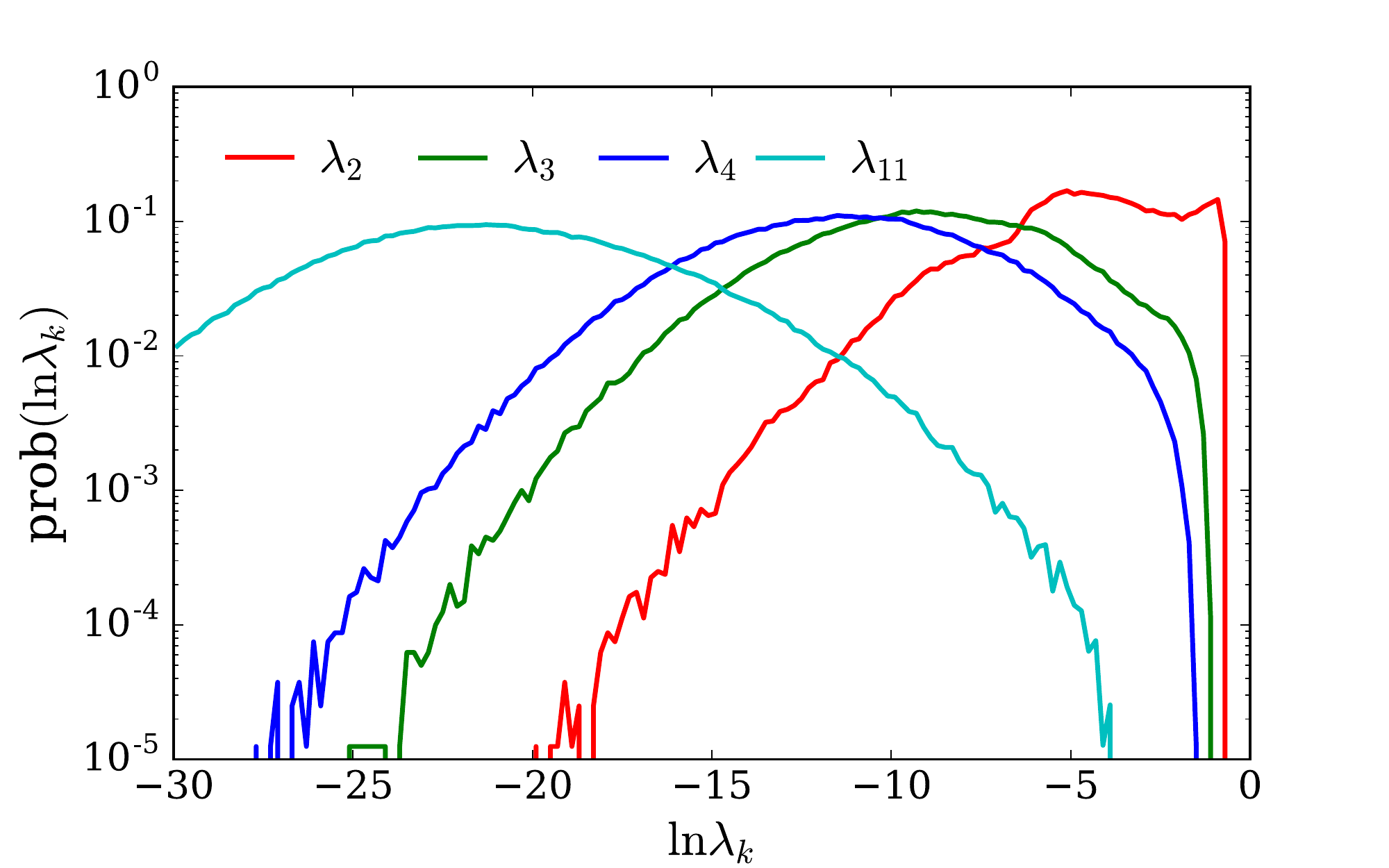}
\caption{ \label{Fig:entspec-stat} All eigenvalues of the density matrix, except for the first few ones, have approximately log-normal distribution for strong disorder (here $W=6.5$ and $L=14$) which is manifested as parabola with on a double logarithmic plot.  }
\end{center}
\end{figure}

Let us illustrate this perturbative process, using just two first blocks, corresponding to $r=0$ and $r=1$:
\begin{equation}\label{Eq:rho1}
 \hat \rho  
 =   
 \ins{\tilde \psi_{1}}\out{\tilde \psi_{1}}
 +
a^2  \ins{\tilde\psi_{2}}\out{\tilde\psi_{2}}
.
\end{equation}
where we define vectors $\ins{\tilde\psi_{1,2}}$ to coincide with $\ins{\psi_{1,2}}$ modulus corresponding factors $a$ that are written explicitly here. Then these vectors have the form: 
\begin{eqnarray}\label{Eq:psi12}
  \ins{\tilde\psi_{1}} &=&  (\alpha_{11},\alpha_{12} a )^T,
  \\
   \ins{\tilde \psi_{2}} &=& (\alpha_{21},\alpha_{22} a )^T,
\end{eqnarray}
where extra factor of $a$ that was present in $\ins{\psi_{2}} $ does not appear in $\ins{\tilde \psi_{2}} $,  $\alpha_{ij}$ denote random numbers which are order one. 

To the leading order in $a$ only the first vector $ \ins{\psi_{1}}$ gives us a non-zero eigenvalue,  $\lambda_1 = \tilde \lambda_1 = O(1)$.  Let us account for the vector $ \ins{\psi_{2}}$ perturbatively. Assuming that vectors $\ins{\tilde\psi_{1}}$ and $\ins{\tilde\psi_{2}}$ are non-orthogonal, we define the vector $\ins{\tilde \psi^{(1)}_2}$ which is orthogonal to $\ins{\tilde\psi_1}$:
\begin{equation}\label{Eq:psi2-ort}
\ins{\tilde\psi^{(1)}_{2}}
  =
\ins{\tilde\psi_{2}}
  -
 \frac{n_{12}}{\tilde \lambda_1} \ins{\tilde\psi_{1}},
\end{equation}
where $n_{12} = \langle \tilde\psi_{1} \ins{\tilde\psi_{2}}\neq 0$ is the overlap between the original vectors, and $\tilde \lambda_1$ is the norm of $\ins{\psi_1}=\ins{\tilde\psi_1}$. Note that while norm of $\ins{\tilde\psi_{2}}$ is order one, the norm of $\ins{\tilde\psi^{(1)}_{2}}$,
\begin{eqnarray}\label{Eq:norm2}
 \langle \tilde\psi^{(1)}_{2} \ins{\tilde\psi^{(1)}_{2}}
 \propto a^2
\end{eqnarray}
becomes suppressed as a result of the orthogonalization process.

Expressing the vector $\ins{\tilde\psi_{2}}$ via $\ins{\tilde\psi^{(1)}_{2}}$, we get:
\begin{multline}\label{Eq:rho1-1}
\hat \rho  
=   
 \left(1 +a^2\frac{n^2_{12}}{\tilde\lambda_1^2}\right)
 \ins{\tilde\psi_{1}}\out{\tilde\psi_{1}}
+
a^2 \ins{\tilde \psi^{(1)}_{2}}\out{\tilde \psi^{(1)}_{2}}
 \\
+
a^2\frac{n_{12}}{\tilde\lambda_1} \left(
\ins{\tilde \psi^{(1)}_{2}}\out{\tilde \psi_{1}}+
\ins{\tilde \psi_{1}}\out{\tilde \psi^{(1)}_{2}}\right),
\end{multline}
where the second line contains off-diagonal terms. These terms can be eliminated using first-order perturbation theory. As a result we get $\lambda_{1} = \tilde \lambda_{1} + O(a^2) =  O(1)$. The next eigenvalue $\lambda_2$ becomes equal to $a^2\corr{\tilde \psi^{(1)}_{2}|\tilde \psi^{(1)}_{2}}\propto a^4$, plus corrections from the perturbation theory which are of the same order. 

Hence, we see that while leading eigenvalue $\lambda_{1} = O(1)$ is still order one, the next eigenvalue is suppressed as 
\begin{eqnarray}\label{Eq:}
\lambda_2\propto a^4
\end{eqnarray}
 The additional  factor of $a^2$ compared to Eq.~(\ref{Eq:r1}) supports the intuition that processes contributing to the entanglement involve similar number of spin flips in the both subsystems, $\cal L$ and $\cal R$. 
  
 \begin{figure}
\begin{center}
\includegraphics[width=0.9\columnwidth]{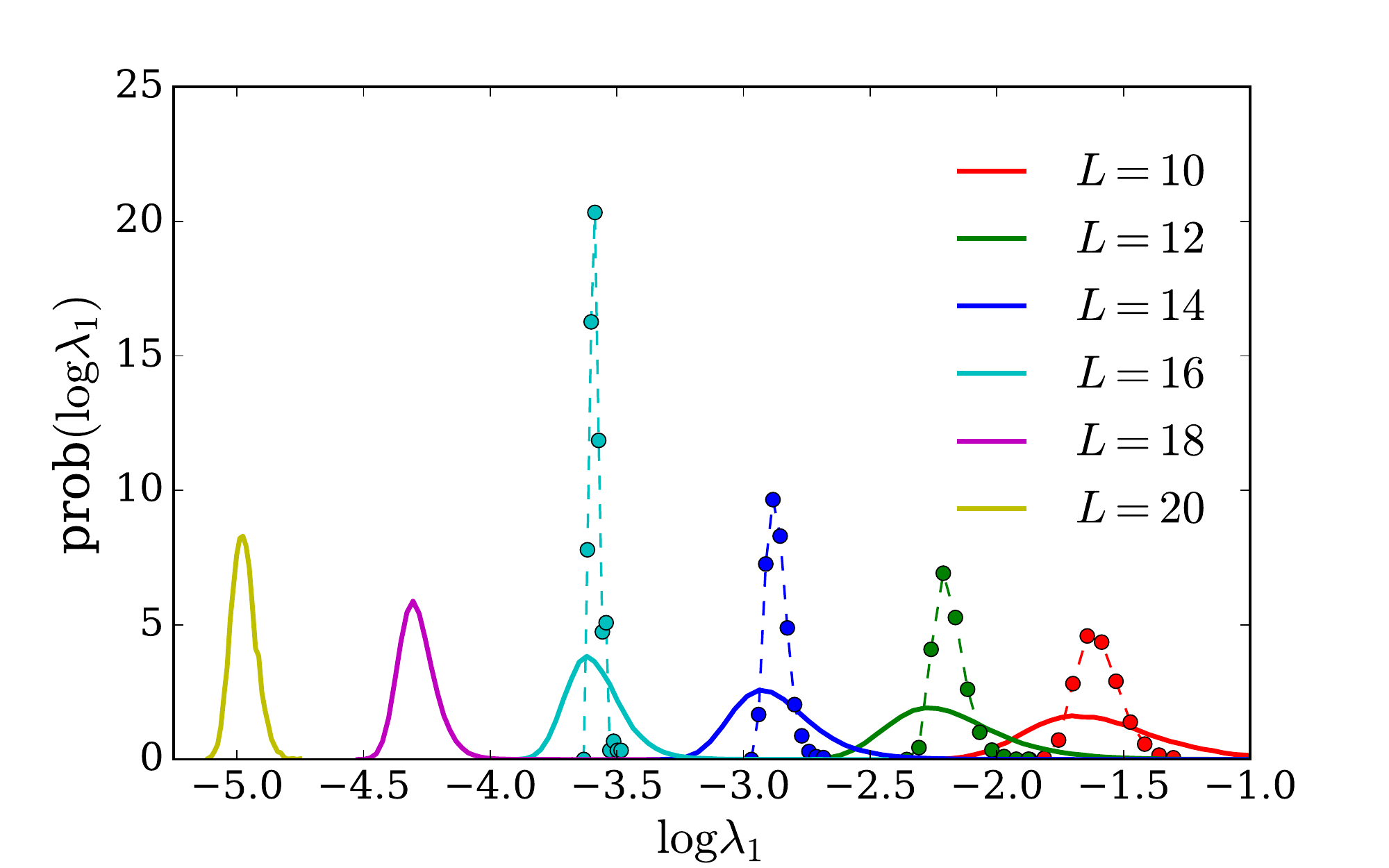}
\caption{ \label{Fig:lmax} Distribution of $\lambda_1$ in the ergodic phase ($W=0.5$) for different $L$. Solid lines indicate full distribution, dashed lines show distribution of $\lambda_1$ between \emph{different disorder realizations}. }
\end{center}
\end{figure} 
  
We illustrated the iterative diagonalization for the first ``block'', involving the $r=0$ and $r=1$ sectors. It can be continued further for $r\geq 2$ sectors. In this process, if we diagonalized all blocks up to $r=r_0$, and are working on the block with $r=r_0+1$, orthogonalization of corresponding vectors $\ins{\tilde\psi_{k}}$ where $k=2^{r_0}+1,\ldots, 2^{r_0+1}$ with all vectors $\ins{\tilde\psi^{(r_0)}_{i}}$ with $i\leq 2^{r_0}$ will lead to new vectors  $\ins{\tilde\psi^{(r_0+1)}_{k}}$ such that $\langle \tilde\psi^{(r_0+1)}_{k}\ins{\tilde\psi^{(r_0+1)}_{k}} \propto a^{2(r_0+1)}$. Hence, we conclude that the eigenvalues of the density matrix will have the hierarchy
\begin{subequations}\label{Eq:hier-true}
\begin{eqnarray}
 r=0: &\qquad\qquad&  \lambda_1  \propto 1,
 \\\label{Eq:r1-true}
 r=1: &\qquad&  \lambda_2 \propto a^4,
  \\
 r=2: &\qquad&   \lambda_{3,4} \propto a^8, 
  \\
   &\ldots&
  \\
  r=L_{\cal L}: &\qquad&     \lambda_{2^{r-1}+1,\ldots, 2^r}\propto a^{4r},
\end{eqnarray}
  \end{subequations}
as we mentioned in the main text.

\subsection{A.2 Tails of the entanglement spectrum}

In the main text we argued that the tails of the entanglement spectrum are described by the so-called ``order statistics'' of the Gaussian distribution.

In the main text, we discussed the distribution of the leading Schmidt eigevalue $\lambda_1$ at strong disorder ($W=6.5$) and near the delocalization transition ($W=2$). In the ergodic phase, Fig. \ref{Fig:lmax} for $W=0.5$,  the center of the distribution of $\lambda_1$ shifts to smaller values according to Marchenko-Pastur formula, and becomes increasingly narrower with increasing $L$. This reflects the fact that all eigenstates become typical (similar to random vectors in the Hilbert space).

\begin{figure}[t]
\begin{center}
\includegraphics[width=0.99\columnwidth]{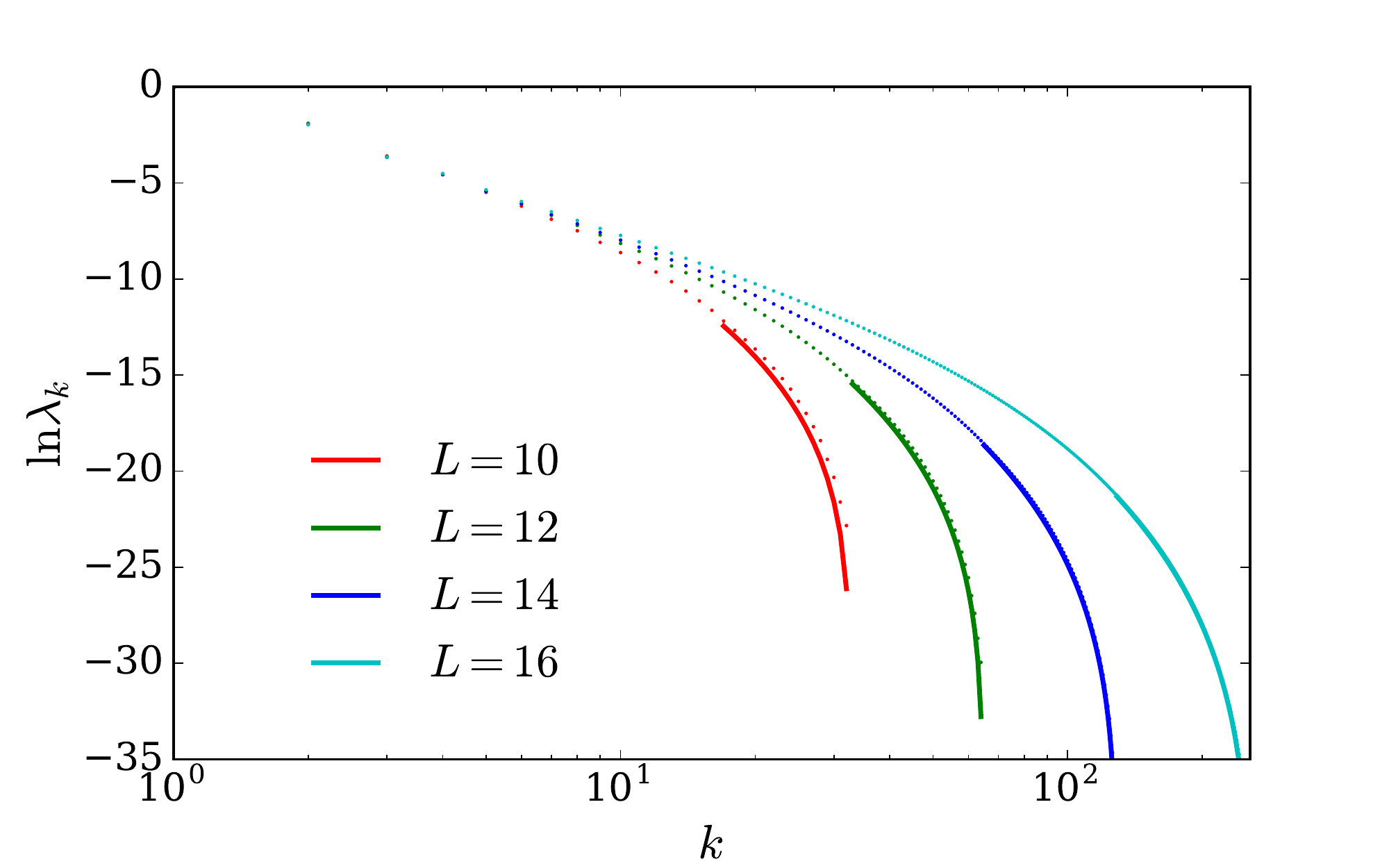}
\caption{ \label{Fig:entspec-tail} Fitting the tails of entanglement spectrum for disorder $W=5$ with the prediction Eq.~(\ref{Eq:tails}). The lines represent the fits, and points are the numerical data.}
\end{center}
\end{figure}

\begin{figure}[b]
\begin{center}
\includegraphics[width=0.99\columnwidth]{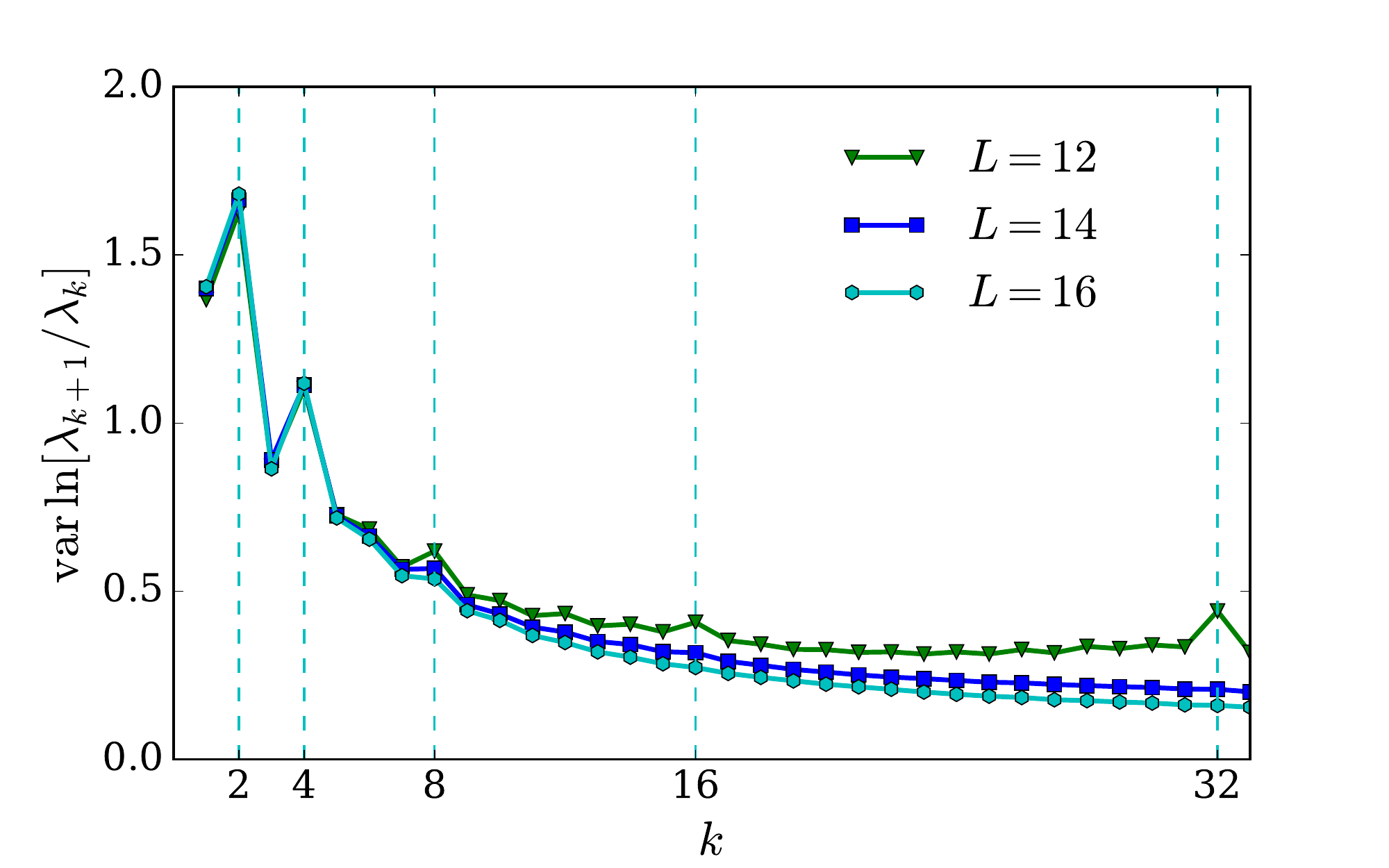}
\caption{ \label{Fig:entspec-jumps} The jumps in the variance of the ratio of different entanglement values illustrate the hierarchy of $\lambda$ in agreement with Eq.~(\ref{Eq:hier-true}). Jumps occur exactly when $k$ is integer power of 2, and get less pronounced for weaker disorder/larger system sizes. Value of disorder here is $W=7$.}
\end{center}
\end{figure}

Next, we study the distribution of the subleading Schmidt eigenvalues. As Fig.~\ref{Fig:entspec-stat} shows, the distribution of $\ln\lambda_k$ for $k\geq 3$ has approximately normal form. Hence, the tails of the ES have to be described by the order statistics for the Gaussian distribution.  The exact analytic form of this statistics is not available, however, it can be approximated by the inverse error function~\cite{Order-stat}. This leads to the following behavior of the tails: 
\begin{equation}\label{Eq:tails}
  \ln \lambda_k
  \approx c_1
  -c_2 \mathop{\rm erf^{-1}}\left[
  \frac{2k-N_{\cal R}-1}{N_{\cal R}-2a+1}
  \right]
  ,
  \quad 
  k\gtrsim 2^{L_{\cal R}-1},
\end{equation}
where $a=0.375$ is a numerical coefficient, $N_{\cal R} = 2^{L_{\cal R}}$, and coefficients $c_{1,2}$ depend on the parameters of the distribution of coefficients, Eq.~(4) in the main text.

Fig.~\ref{Fig:entspec-tail} demonstrates that tails of the ES can be well approximated by the Gaussian order statistics~\footnote{We note, that while we plot  $\{\lambda_k\}$, we use direct SVD algorithm in our numerics. Such algorithm gives us singular values $\{\sqrt{\lambda_k}\}$ with double precision, hence potentially allowing us to resolve $\lambda_k$ as small as $\sim 10^{-30}\approx e^{-70}$.}. The solid lines in Fig.~\ref{Fig:entspec-tail} correspond to the Eq.~(\ref{Eq:tails}), where coefficients $c_{1,2}$ were determined from matching the first and last data point in the ES tail. 

Finally, to further support the existence of hierarchical sectors in the ES, we study the variance of the ``entanglement gap'', being defined as $\ln \lambda_{k+1} - \ln \lambda_k$. Figure~\ref{Fig:entspec-jumps} illustrates that entanglement gap has much broader distribution when $k=2,4,8,16,\ldots$. These are the exact points where one goes between blocks labelled by different values of $r$ in the hierarchy of the ES, see Eq.~(\ref{Eq:hier-true}).

\subsection{A.3 Entanglement spectrum of rare many-body resonances}

Here we address the entanglement spectrum in eigenstates which contain (rare)  finite regions of size $\ell$ that (locally) have volume-law entanglement. In order to have a stable MBL phase, the probability to find such regions is expected to be $P(\ell)=e^{-\alpha\ell}$ (this is equivalent to the exponential decay of the entanglement entropy probability distribution, see Fig. 9 in Ref.~\cite{Clark16}). Then we can generalize Eq.~(4) of the main text  as follows. 

For simplicity, let us take an eigenstate where the middle (``hot") region of size $\ell$ has volume-law entropy, and assume that we measure the ES around the center of that region. The ES in an area of radius $\ell/2$ around the cut may be complicated; if the region is truly thermal, we would expect a very flat distribution of the ES eigenvalues according to Marchenko-Pastur. More importantly, Eq.~(4) holds for the sites outside that area, for the same reason as we have described in the main text. In terms of the Hilbert space, the first $d^{\ell/2}$ reduced density matrix eigenvalues ($d$ is local Hilbert space dimension, i.e., $d=2$ for spin-1/2 models) have a structure dictated by the finite hot region, while for the rest of the spectrum farther away we expect a power law. We have to note that the bound $d^{\ell/2}$ is not strict and a few sites around the boundary may exhibit an intermediate behavior due to the coupling to the hot region.

\begin{figure*}[t]
\centering 
\includegraphics[width=0.95\textwidth]{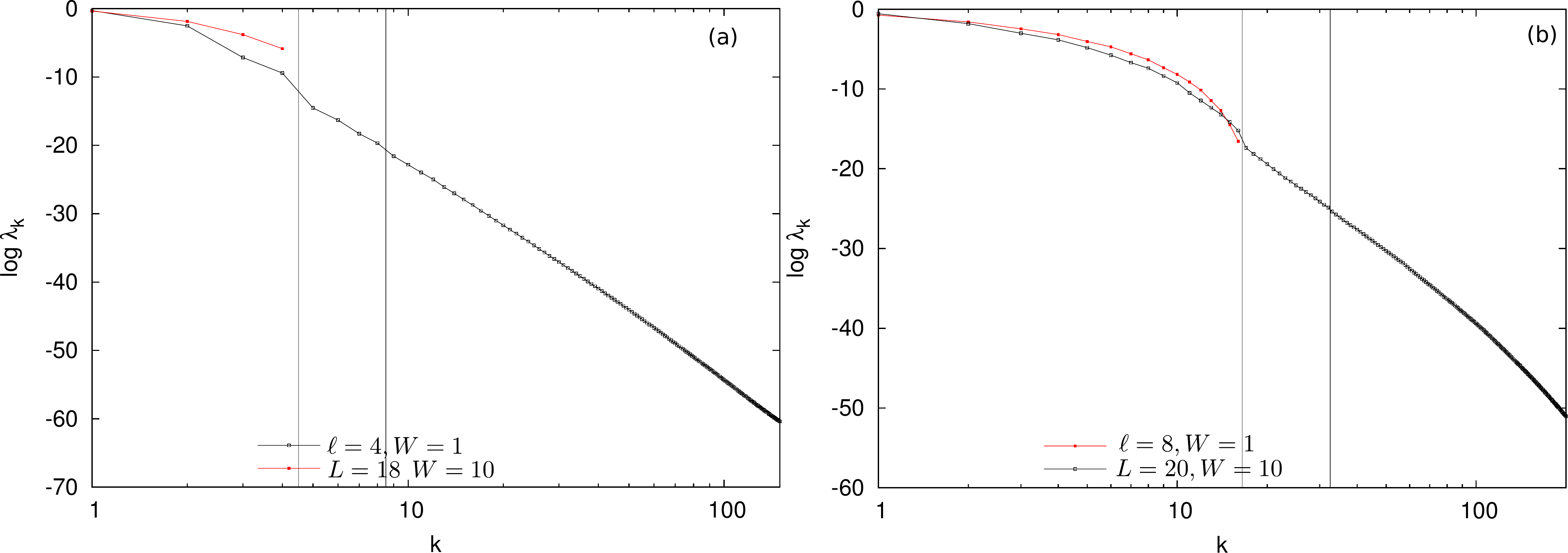}
\caption{ \label{fig:res} The averaged typical entanglement spectrum for a system of size $L$ containing a middle hot region of size $\ell$. (a) Total chain size is $L=18$ with disorder $W=10$, and a hot region of size $\ell=4$ and disorder $W=1$. (b) Total chain size of $L=20$ with disorder $W=10$, and a hot region of size $\ell=8$ and disorder $W=1$. Red lines correspond to the ES of a system of size $\ell$ and disorder $W=1$. Vertical lines are guide to the eye and denote $d^{\ell/2}$ and $d^{1+\ell/2}$.}
\end{figure*}

To support the simple picture from the previous paragraph, we created artificial many-body resonances by including a slightly disordered region (with disorder strength $W=1$) coupled to a highly disordered ($W=10$) bath. In Fig. \ref{fig:res} we computed the averaged entanglement spectrum for two such configurations: (a) total chain size of $L=18$ with disorder $W=10$, and a hot region of size $\ell=4$ and disorder $W=1$; and (b) total chain size of $L=20$ with disorder $W=10$, and a hot region of size $\ell=8$ and disorder $W=1$. In both cases, we first notice a strong kink located at precisely $d^{\ell/2}$. This separates the ``thermal" part of the ES from the rest, as we can show by comparing this ES to the ES of a full system containing $\ell$ spins at disorder $W=1$. When the hot region is small [Fig.  \ref{fig:res}(a)] there is some discrepancy between the two, but as the hot region becomes larger, its ES approaches the ES of a thermal system [Fig. \ref{fig:res}(b)]. 

At large distances from the ES cut, $k\gg d^{\ell/2}$, we expect the power law to kick in. However, the first few spins neighboring the hot region may be strongly affected, and thus their ES may deviate from the power law. This means we might expect smaller kinks in the ES for $k=d^{\ell/2+1}, d^{\ell/2+2}, \ldots$, which should terminate once the power law sets in. In Fig. \ref{fig:res} we see that this happens very rapidly, i.e., roughly after only one additional spin adjacent to the cut. Indeed, there is a very small kink at $k=d^{1+\ell/2}$, and beyond this point we obtain the perfect power law. For values $d^{\ell/2}<k<d^{\ell/2+1}$, the ES possibly has a complicated form that is an intermediate between thermal and power-law.   
\begin{figure}[b]
\begin{center}
\includegraphics[width=0.99\columnwidth]{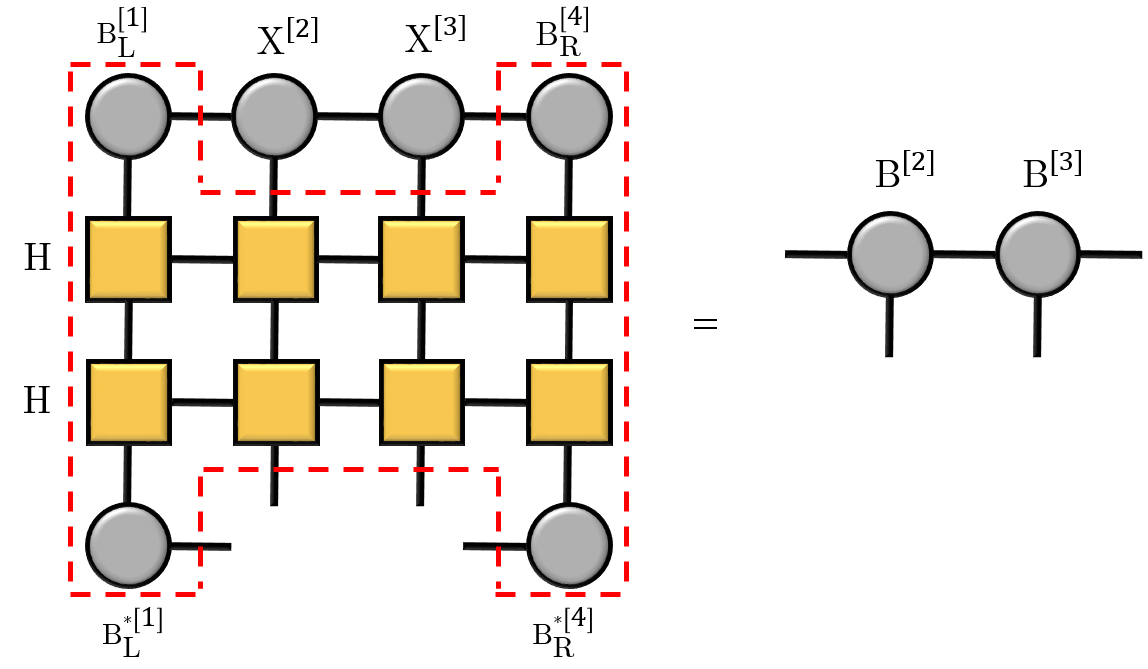}
\caption{ \label{Fig:MPO_MPS} Equation (\ref{Linear_eq}) in graphical form. Tensors $X$ are the solutions of the system. $L/R$ denote left/right orthonormal operators. The red boundary denotes $(\tilde{H}_{eff}^{[2,3]})^2$. On the right hand side, the rest of the network contract to identities as a result of Eq. (\ref{Left/Right}).}
\end{center}
\end{figure}

\section{B. MPO Inversion using Matrix Product States}

\subsection{B.1 Details of the algorithm}

When we aim to target an eigenvalue $E$ of a matrix $H$ at a specific part of the spectrum efficiently, we shift-invert the matrix 
\begin{equation}\label{Eq:tildeH}
\tilde{H}^{-1} = (H- E I)^{-1}.
\end{equation}
The ground state of $\tilde{H}^{-1}$ is the target state and a power method can be employed to calculate it efficiently. Direct inversion of an MPO version of $\tilde{H}$ has an MPO dimension $D \propto 2^{L}$ as it contains arbitrary many-body long-range terms, rendering it inefficient. Thus, following the recipe of two-site DMRG algorithm, we have implemented an algorithm which consists of global and local iterations.

In the global iteration we use power method
\begin{equation}\label{power_method}
\begin{aligned}
&\ins{\psi}_{i+1} = (\tilde{H}^{-1})^2\ins{\psi}_{i} \\
&\ins{\psi}_{i} = \tilde{H}^2\ins{\psi}_{i+1}
\end{aligned}
\end{equation}
where $i$ is the iteration index. Instead of multiplying a vector by the inverse matrix $(\tilde{H}^{-1})^2$, we seek the solution of a linear system. The operator $\tilde{H}$ is squared to make it sign-definite. This improves the stability and allows to use the conjugate gradient (CG) algorithm to solve the linear system. In MPS notation, one iteration of the power method corresponds to a sweep of the chain. Of course, since the iteration is broken in local pieces, the number of power iterations needed to target a state precisely is much higher than a shift-invert power method operating in the full Hilbert space.
The MPO representation of $\tilde{H}$ can be squared locally to a MPO of bond dimension $D^2$ or a compressed version of dimension $2D-1$. Squaring changes trivially the spectrum when the matrix is Hermitian and no degeneracies to the spectrum are formed.

\begin{figure}[t]

\begin{overpic}[width=0.9\columnwidth,tics=10]{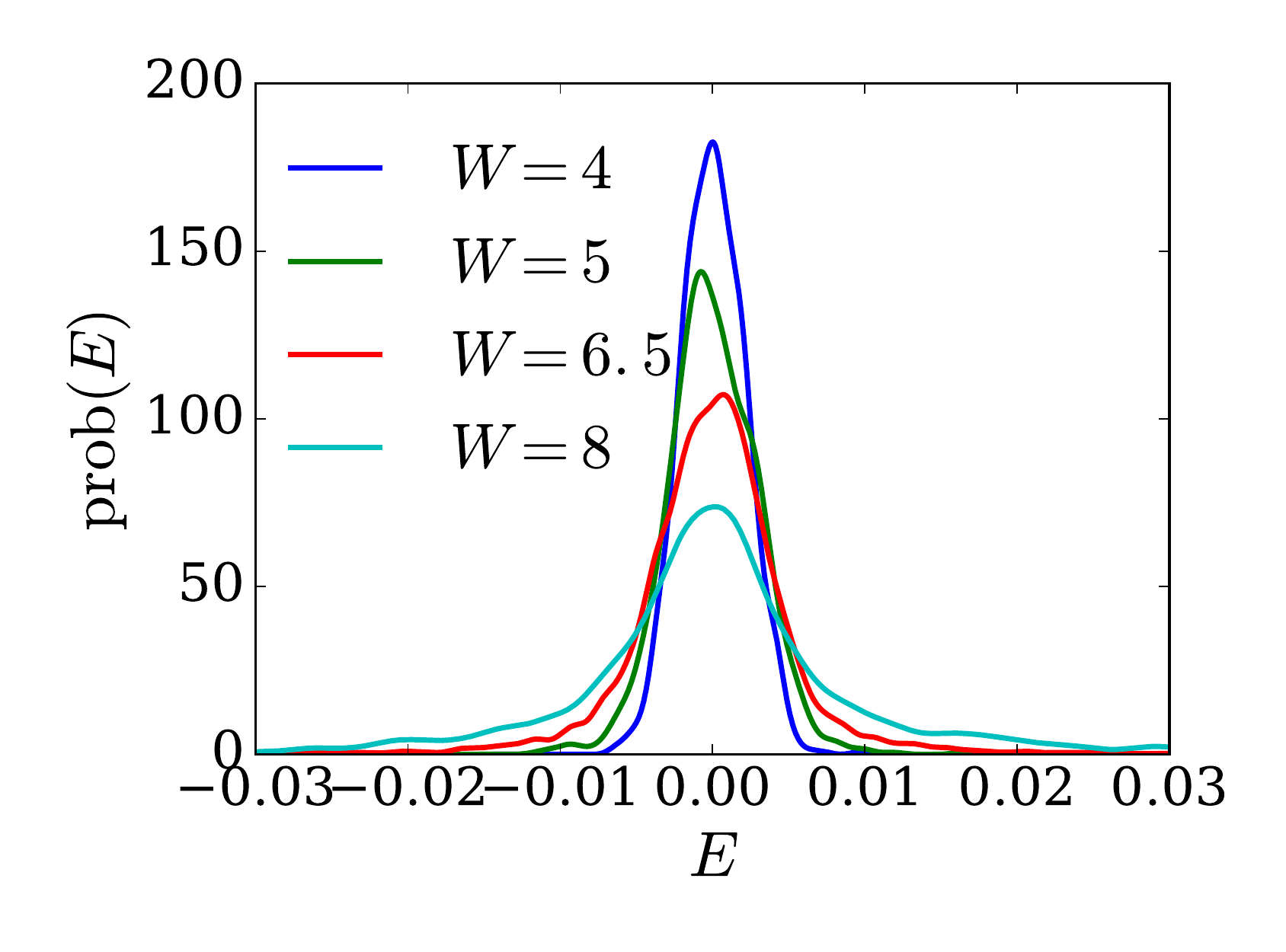}
 \put (0,70) {(a)}
\end{overpic}

\begin{overpic}[width=0.9\columnwidth,tics=10]{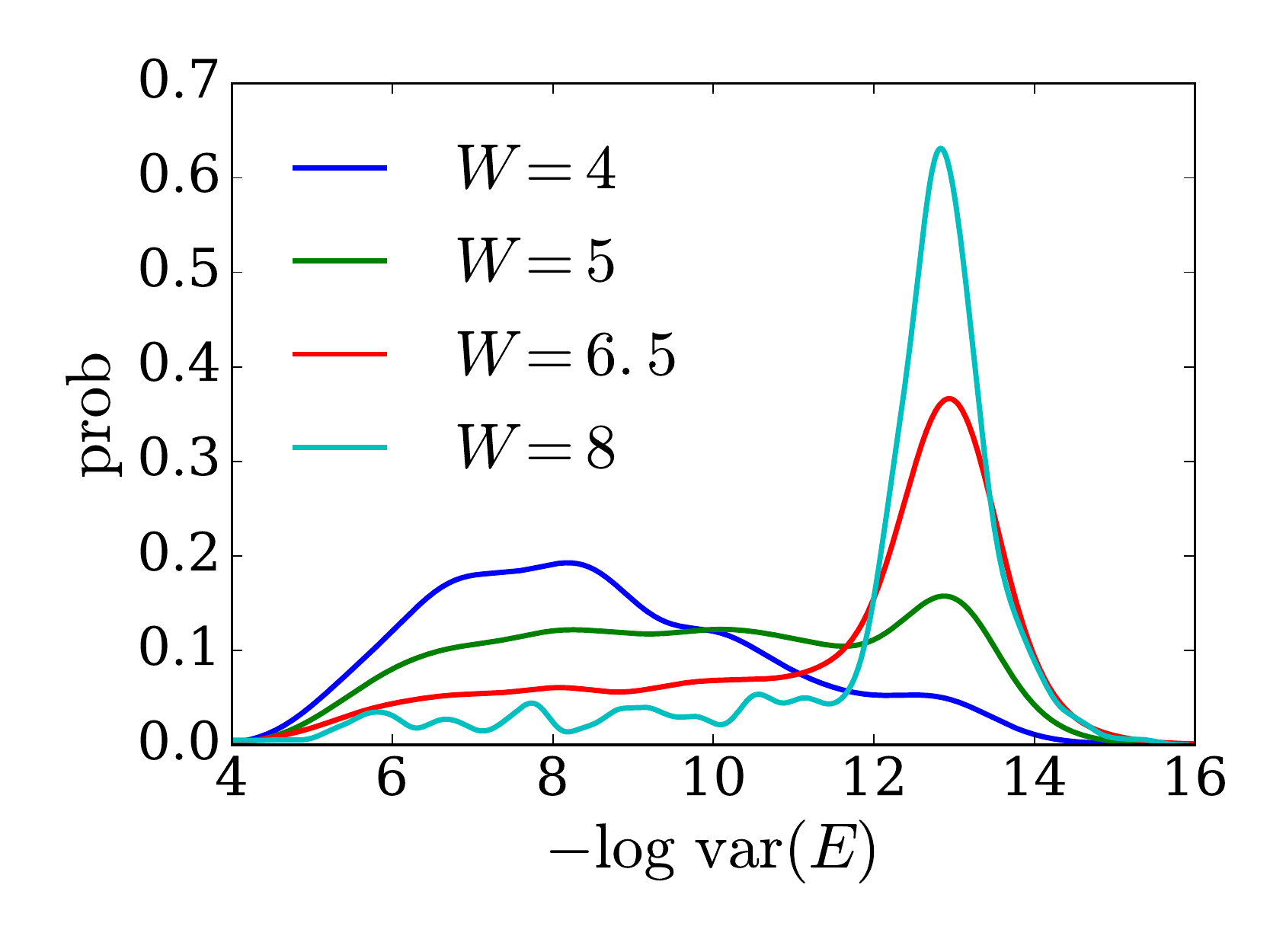}
 \put (0,70) {(b)}
\end{overpic}

\caption{ \label{Fig:probE} Probability density functions of energy (a) and logarithm of energy variance (b) for various disorder amplitudes. The peak in (b) around $13$ corresponds to states which converged to machine precision.
}
\end{figure}

\begin{figure}[t]

\begin{overpic}[width=0.9\columnwidth,tics=10]{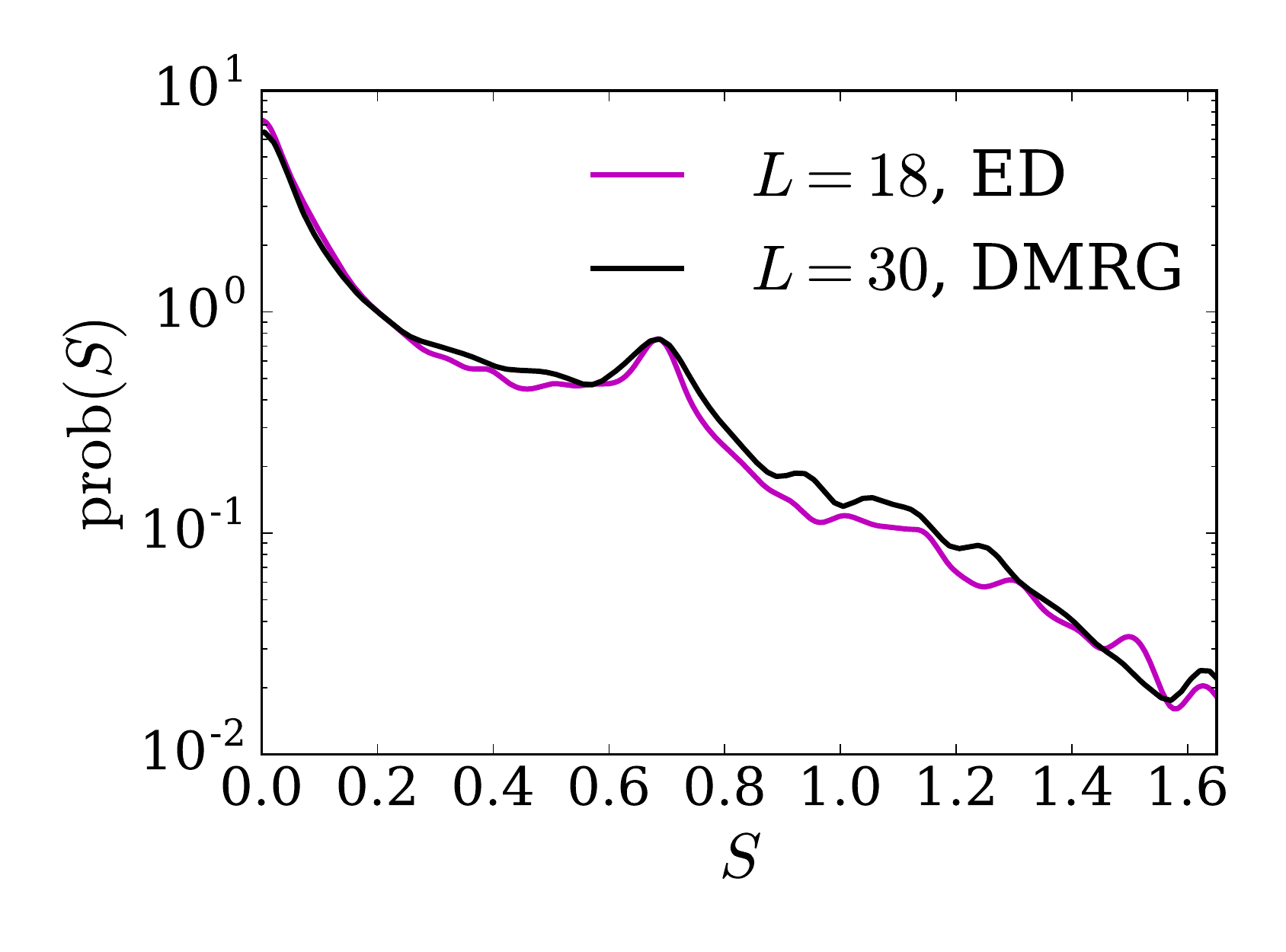}
 \put (0,70) {(a)}
\end{overpic}

\begin{overpic}[width=0.9\columnwidth,tics=10]{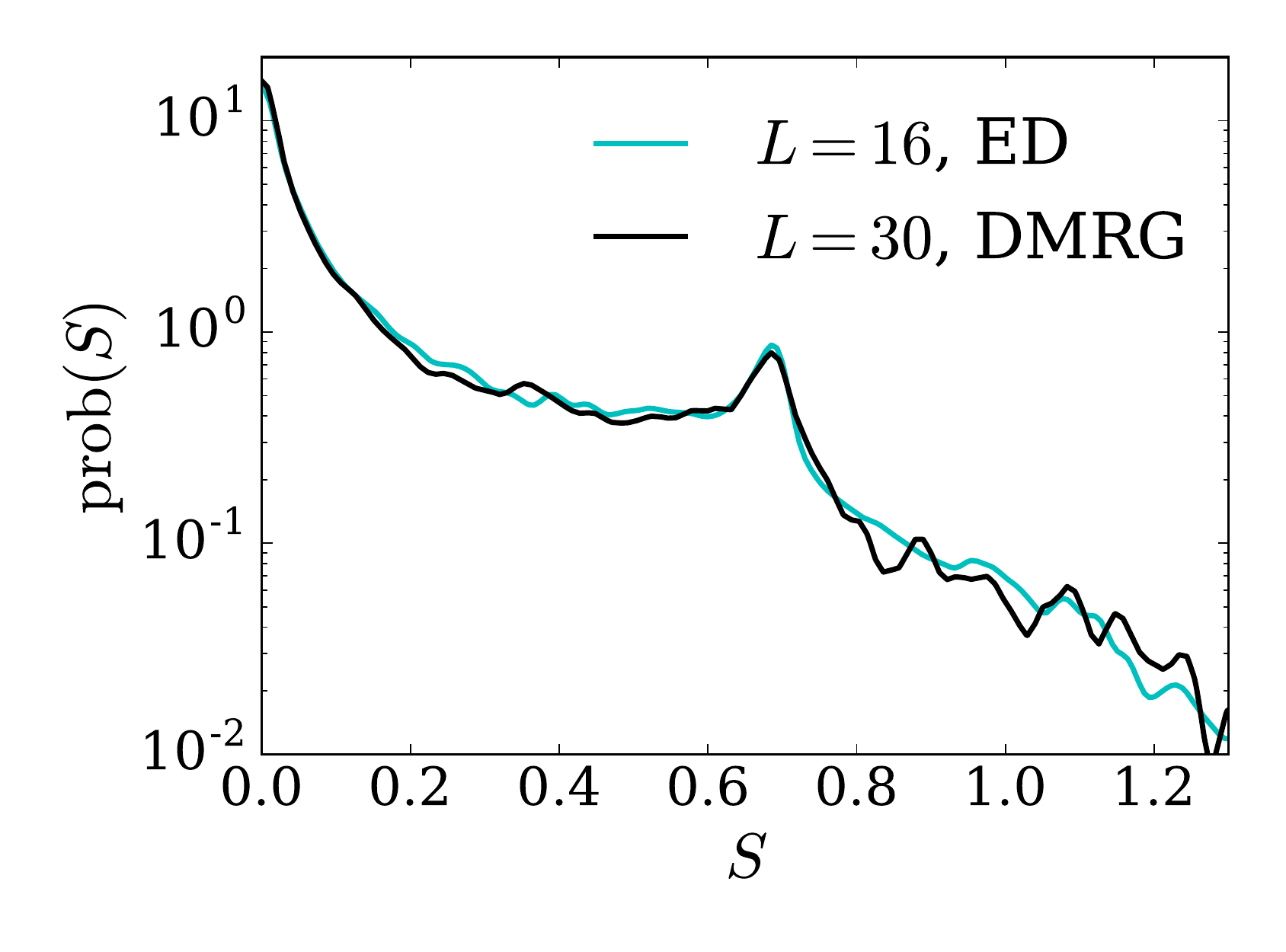}
 \put (0,70) {(b)}
\end{overpic}

\caption{ \label{Fig:probS} Probability density functions of typical entanglement entropy at disorder $W=5$ (a) and $W=6.5$ (b). The samples include 4000 disorder realizations. We measure entanglement in 3 central bonds per state. 
}
\end{figure}

Each global iteration consists of $L-1$ local iterations. In a local iteration, we decompose the power method in a series of two-site optimization operations (DMRG-style). For each optimization the MPS is prepared in mixed gauge form 
\begin{multline}\label{state}
\ins{\psi} = \text{Tr} \{ B_{\alpha_{j-1}\alpha_{j}}^{[j]s_j}B_{\alpha_{j}\alpha_{j+1}}^{[j+1]s_{j+1}}\\
\times\ins{\alpha_{j-1} L_{j-1}}\ins{s_{j}s_{j+1}} \ins{\alpha_{j+1} R_{j+1}} \},
\end{multline}
where trace sums over all virtual and site indices, and 
\begin{equation}\label{Left/Right}
\begin{aligned}
&\langle{\alpha_{j-1} L_{j-1}}\ins{\alpha_{j-1}' L_{j-1}}=\delta_{\alpha_{j-1}\alpha_{j-1}'}\\
&\langle{\alpha_{j+1} R_{j+1}}\ins{\alpha_{j+1}' R_{j+1}}=\delta_{\alpha_{j+1}\alpha_{j+1}'}.
\end{aligned}
\end{equation}
We want to optimize sites $j,j+1$. So, we propose a variational state 
\begin{multline}
\ins{\theta} = \text{Tr} \{ X_{\alpha_{j-1}\alpha_{j}}^{[j]s_j}X_{\alpha_{j}\alpha_{j+1}}^{[j+1]s_{j+1}}\\
\times\ins{\alpha_{j-1} L_{j-1}}\ins{s_{j}s_{j+1}} \ins{\alpha_{j+1} R_{j+1}} \},
\end{multline}
by replacing the known tensors $B$ in $\ins{\psi}$ with unknown tensors $X$ only in the sites $j,j+1$. 
We then solve the linear system
\begin{equation}\label{Linear_eq-0}
\text{Tr} \{ \tilde{H}^2\ins{\theta} \} = \text{Tr} \{ \ins{\psi} \},
\end{equation} 
where trace denotes the inner product with the left and right part of the state as defined in Eq. (\ref{state}).
Using Eq. (\ref{Left/Right}), for each two-site operation we need to solve a linear system
\begin{equation}\label{Linear_eq}
(\tilde{H}_{eff}^{[j,j+1]})^2X^{[j,j+1]} = B^{[j,j+1]},
\end{equation}
where the compact notation is explained in Fig. (\ref{Fig:MPO_MPS}).
 We solve the local linear system using a CG algorithm. In the CG algorithm we use Polak-Ribi\`ere formula \cite{PolakRibiere} for the parameter $\beta$ which controls the direction of descend, as we found it more efficient than other variants. We make $N_{CG}$ iterations of CG algorithm for each two-site update, normalize the vector and continue to the next update. This corresponds to one local iteration of the power method.

In the simulations we keep the number of global iterations $N_{G}=50$ and the number of local iterations $N_{L}=1$. For each local iteration we make $N_{CG}=250$ CG iterations. We can replace the CG algorithm with an exact inversion scheme at the cost of $\sim O(\chi^2/N_{CG})$, which restricts the applicability of exact inversion to very small bond dimensions $\chi$. We also used the U(1) symmetry of the XXZ model to allow for larger dimension $\chi$. The starting states are random product states at half-filling. We do not filter any states to avoid influencing the probability distributions. The bond dimension used is $\chi_{max}=400$ for $W=4$ and $\chi_{max}=200$ for the rest. Dimension $\chi$ is adaptive at each bond with a cutoff at machine precision magnitude of the Schmidt coefficients. The allowed maximal entanglement $ \approx \ln(\chi_{max})$  is much higher than the maximum entanglement measured in the samples. We target zero energy, $E=0$, which approximately corresponds to the middle of the many-body band. The size of the chain is fixed to $L=30$.

\subsection{B.2 Performance of the algorithm and discussion}

There are two types of error in our MPS algorithm. The first one is the ``distance'' of the converged state from the exact eigenstate of the Hamiltonian, which can be measured by energy variance, $\langle H^2\rangle-\langle H \rangle^2$. The second error comes from the fact that we deal with a dense (beyond machine precision) spectrum. Higher disorder corresponds to steeper energy landscape, which in turn means that the algorithm may ``get stuck" far from the target energy. On the other hand, as hybridization increases for lower disorder, we expect to get a sharper distribution and an increase of difficulty to distinguish between states, which should result in higher variance. Such effects can be observed in Fig. \ref{Fig:probE}.

We further quantify the degree to which the energy variance affects the probability distribution of the entanglement entropy. We find in Fig.(\ref{Fig:probS}) that for $W=5$ and $W=6.5$, the accessible part of the distribution is not visibly affected. In particular, we find that the tail of the distribution $P(S)$ agrees accurately with ED results, thus there is no issue with the bias towards low-entangled states, which could be a potential problem with the DMRG.

\begin{figure}
\begin{center}
\includegraphics[width=0.96\columnwidth]{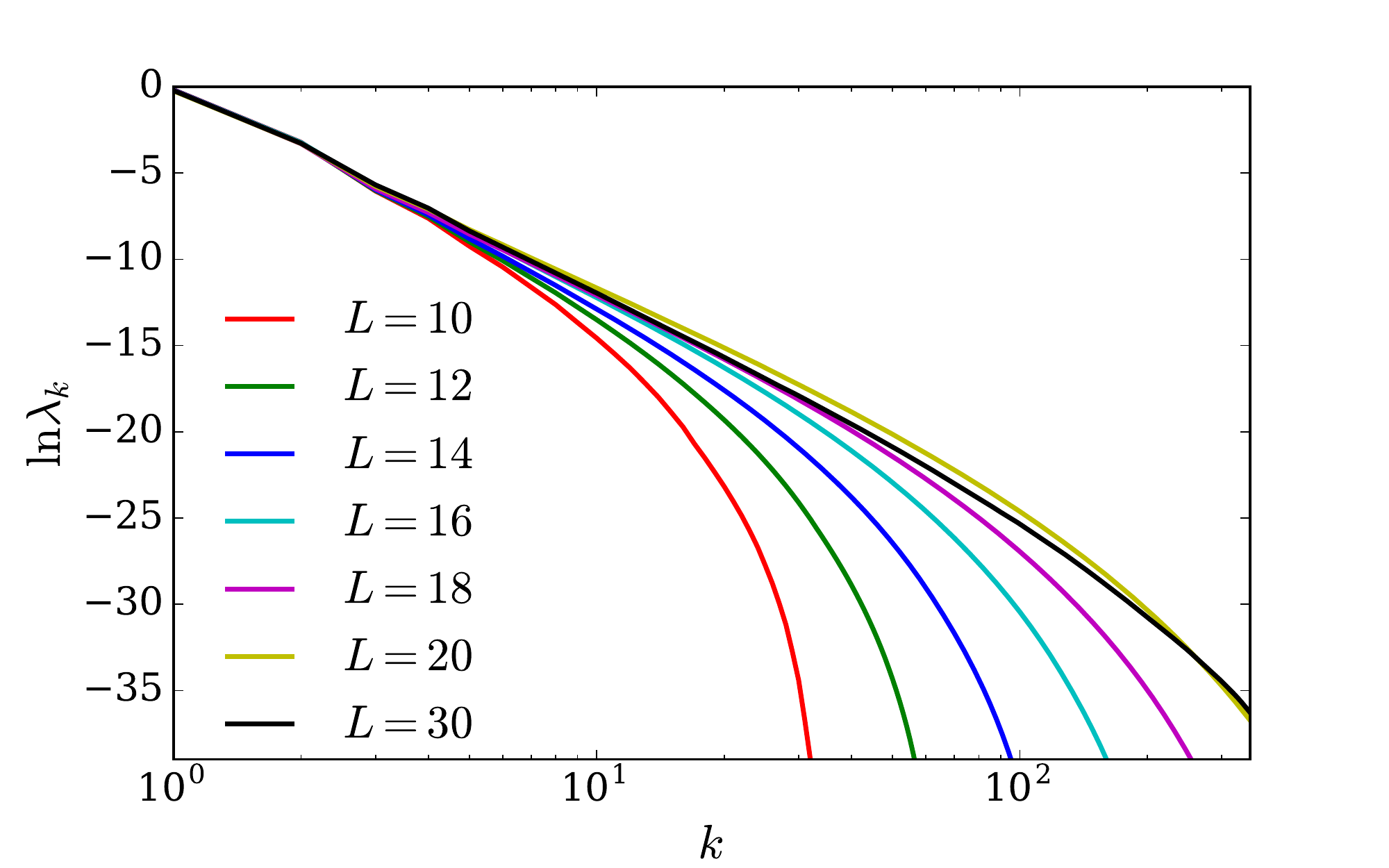}
\caption{ \label{Fig:entspecW4} Comparing data from MPS algorithm with the data from ED and SI for the entanglement spectrum for the XXZ spin chain with disorder $W=4$ near the MBL transition. }
\end{center}
\end{figure}

Finally, we have pushed our calculations to disorder $W=4$, which places the system very near the MBL transition. In this case, we find a broader distribution of energy variance (Fig. \ref{Fig:probE}), and one may worry that this would strongly affect the slope of the ES, which is determined by smaller Schmidt values. As we observe in Fig.(\ref{Fig:entspecW4}), the relatively large energy variance at disorder $W=4$ is not enough to produce inaccuracies in the typical entanglement spectrum. Comparison of the spectrum obtained by MPS even in this case agrees remarkably accurately with the one obtained by ED.

An important strength of our iterative scheme is that for low disorder, e.g., $W=4$, we observed several bonds with entanglement entropy $S>3$. This would render the algorithms with $\ln(\chi) \approx S$ (the range where exact inversion is applicable) biased against such states, which are rare but important. To access those states we believe that our algorithm is the most appropriate up-to-date scheme, as it combines the accuracy of shift-invert technique and the flexibility of the CG scheme. Of course, the CG scheme induces extra errors, but as we have shown they are not severe; for example, our obtained energy variance (Fig. \ref{Fig:probE}) is comparable or lower than other approaches ~\cite{Pekker15,Sheng15,Pollmann15-1,Pollmann15,Karrasch15}. Finally, to reduce the errors even further, it is possible to increase the number of CG iterations $N_{CG}$, which is far more efficient than increasing the bond dimension $\chi$ in methods relying on exact inversion.


\end{document}